\documentclass[pra,letterpaper,aps,10pt,superscriptaddress,twocolumn,floatfix,showpacs]{revtex4-1}
\usepackage{graphicx}
\usepackage{amsmath}
\usepackage{amsfonts}
\usepackage{float}
\usepackage{soul}
\normalsize
\usepackage{booktabs}
\newcommand\identity{1\kern-0.25em\text{l}}
\usepackage{amssymb}
\usepackage{epsfig}
\usepackage{epstopdf}
\DeclareGraphicsExtensions{.pdf,.eps,.png,.jpg,.mps,.heic}
\usepackage[pdftex]{color}
\usepackage{amsmath,graphicx,amssymb,braket,xcolor,subfigure,upgreek}
\usepackage[colorlinks, linkcolor=blue, citecolor=blue, urlcolor=blue, breaklinks=true]{hyperref}
\usepackage{microtype}
\usepackage{bbm}
\usepackage{color}
\usepackage{dsfont}
\usepackage[english]{babel}
\usepackage{blindtext}
\usepackage{textcomp}
\usepackage{svg}
\newcommand{\s}{\sigma}

\newcommand{\calS}{\mathcal{S}}

\newcommand{\td}{\mathrm{d}}

\newcommand{\ts}[1]{_\text{#1}}

\newcommand{\Hm}{\mathcal{H}}

\newcommand{\br}{\mathbf{r}}

\newcommand{\bk}{\mathbf{k}}
\newcommand{\bq}{\mathbf{q}}
\newcommand{\bE}{\mathbf{E}}
\newcommand{\calH}{\mathcal{H}}
\newcommand{\bmu}{\boldsymbol{\mu}}

\newcommand{\beps}{\boldsymbol{\varepsilon}}

\newcommand{\me}{\mathrm{e}}
\newcommand{\mi}{\mathrm{i}}
\newcommand{\iu}{\mathrm{i}} 
\usepackage{physics}

\newcommand{\rev}[1]{\textcolor{black}{#1}}

\newcommand{\revise}[1]{\textcolor{black}{#1}}

\begin{document}

\title{Quantum theory of surface lattice resonances}
\author{M.~Reitz}
\affiliation{Department of Chemistry and Biochemistry, University of California San Diego, La Jolla, California 92093, USA}
\author{S.~v.~d.~Wildenberg}
\author{A.~Koner}
\affiliation{Department of Chemistry and Biochemistry, University of California San Diego, La Jolla, California 92093, USA}
\author{G.~C.~Schatz}
\affiliation{Department of Chemistry, Northwestern University, Evanston, Illinois 60208, USA}
\author{J.~Yuen-Zhou}
\email{joelyuen@ucsd.edu}
\affiliation{Department of Chemistry and Biochemistry, University of California San Diego, La Jolla, California 92093, USA}
\date{\today}

\begin{abstract}
The collective interactions of nanoparticles arranged in periodic structures give rise to high-$Q$ in-plane diffractive modes known as surface lattice resonances. While these resonances and their broader implications have been extensively studied within the framework of classical electrodynamics and linear response theory, a quantum optical theory capable of describing the dynamics of these structures, especially in the presence of material nonlinearities beyond \textit{ad hoc} few-mode approximations, is largely missing. To this end, we consider a lattice of metallic nanoparticles coupled to the electromagnetic field and derive the quantum input--output relations within the electric dipole approximation. As applications, we analyze coupling between the nanoparticle array and external quantum emitters, and show how the formalism extends to molecular optomechanics, where the high $Q$-factors of SLRs enable coupling to collective vibrational modes. We further consider arrays composed of saturable excitonic emitters, demonstrating how emitter nonlinearities can be used to switch the SLR condition between electronic transitions. Using a perturbative approach that accounts for population dynamics, we show how these effects can be probed in pump--probe experiments and give rise to nonlinear phase-matching phenomena. Our work provides a microscopic framework for modeling SLRs interacting with quantum emitters without phenomenological descriptions of the electromagnetic environment.
\end{abstract}

\maketitle


\section{Motivation}

Light scattering from interfaces exhibiting periodic structuring at or below the scale of the incident wavelength produces remarkable optical effects through wave diffraction and interference, enabling precise control over the propagation of light \cite{garcia2007colloquium}. Over the last 20 years, arrays of metallic nanoparticles (MNPs) have been investigated as platforms with great tunability in terms of spatial arrangement, where the localized surface plasmons (LSPs) originating from  collective oscillations of the electrons inside the particles can provide large enhancements of the electromagnetic field amplitude in their vicinity \cite{maier2007plasmonics}. Moreover, interactions among nanoparticles in the lattice can give rise to a coupling between the LSPs and diffractive light modes propagating in the lattice plane, resulting in collective, hybrid photonic-plasmonic modes characterized by extremely narrow lineshapes known as \textit{surface lattice resonances} (SLRs) \cite{zou2004narrow, zou2004silver, auguie2008collective, kravets2008extremely, chu2008experimental, vitrey2014parallel, guo2017geometry, cherqui2019plasmonic}.  Due to their unique and highly tailorable features, these collective resonances have found a wide range of applications, e.g., in sensing \cite{danilov2018ultra, matricardi2018gold, thackray2014narrow}, lasing \cite{wang2018structural, hakala2017lasing, pourjamal2019lasing} and condensation \cite{rodriguez2013thermalization, hakala2018bose, Taskinen2021polarization}, as nanophotonic devices \cite{lozano2013plasmonics, wang2018rich, bourgeois2022optical}, or for the implementation of linear and nonlinear optical elements \cite{hu2019lattice, kataja2015surface, czaplicki2016multiharmonic, binalam2021ultra}. Plasmonic lattices share similarities with quantum optical platforms such as atomic arrays \cite{bettles2016enhanced, shahmoon2017cooperative, rui2020asubradiant, bekenstein2020quantum, masson2020atomic} which are, however, typically treated within the Markovian approximation. 

SLRs and their interactions have been widely described using classical  electromagnetic approaches, including Green's function methods and coupled dipole theory \cite{cherqui2019plasmonic, zundel2022green, Baur2018hybridization}. Moreover, macroscopic quantum electrodynamics (QED) descriptions have been developed to investigate the interaction of quantum emitters with plasmons in single metal nanoparticles—encompassing the strong coupling regime in plasmonic systems \cite{rousseaux2018quantum, rousseaux2020strong, wang2023exploring}. Moreover, theoretical treatments that extend beyond the dipole approximation have been introduced to account for higher multipole contributions \cite{miwa2021quantum}. In this work, we develop a quantum input-output formalism for the description of SLRs which provides a consistent description of both collective plasmonic resonances and the matter counterpart on a common theoretical footing. The hybridization of the LSPs with the radiative continuum emphasizes the open-system nature of the problem and cautions against uncontrolled phenomenological few-mode descriptions.
Since this necessarily invokes a description beyond the Markovian regime, we derive expressions for the dipole amplitudes in Fourier domain, allowing a direct relation to the input electric field.

This paper is organized as follows. We start by deriving the optical response of an array of nanoparticles to a quantized electromagnetic field and illustrate how SLRs emerge due to the coupling between LSPs and diffractive modes of the lattice. This exercise provides results that are consistent with classical coupled-dipole methods, yet starting from a Hamiltonian formalism lends itself naturally for generalizations involving quantum emitters with complex internal structure. In particular, we apply the formalism towards molecular optomechanics and show how the high $Q$-factors exhibited by SLRs facilitate the coupling between lattice resonances and collective molecular vibrational modes. Finally, we consider replacing the nanoparticles with saturable emitters such as molecules, giving rise to excitonic SLRs. We then show how the emitter nonlinearity can be harnessed to switch the SLR condition between distinct electronic states \revise{under the assumption of static populations. We then extend the treatment to include dynamical population effects in a perturbative fashion, demonstrating that the resulting switching dynamics can be observed and characterized in pump–probe experiments.} Our Hamiltonian formalism provides a transparent pathway towards the description of complex material nonlinearities in the optical response of plasmonic arrays and is completely compatible with more sophisticated approaches to treat the plasmonic media (e.g., macroscopic QED \cite{buhmann2007dispersion, siwei2019quantum, feist2021macroscopic}).

\section{Model}
\label{sec:model}

\begin{figure}[t]
	\centering
		\includegraphics[width=0.99\columnwidth]{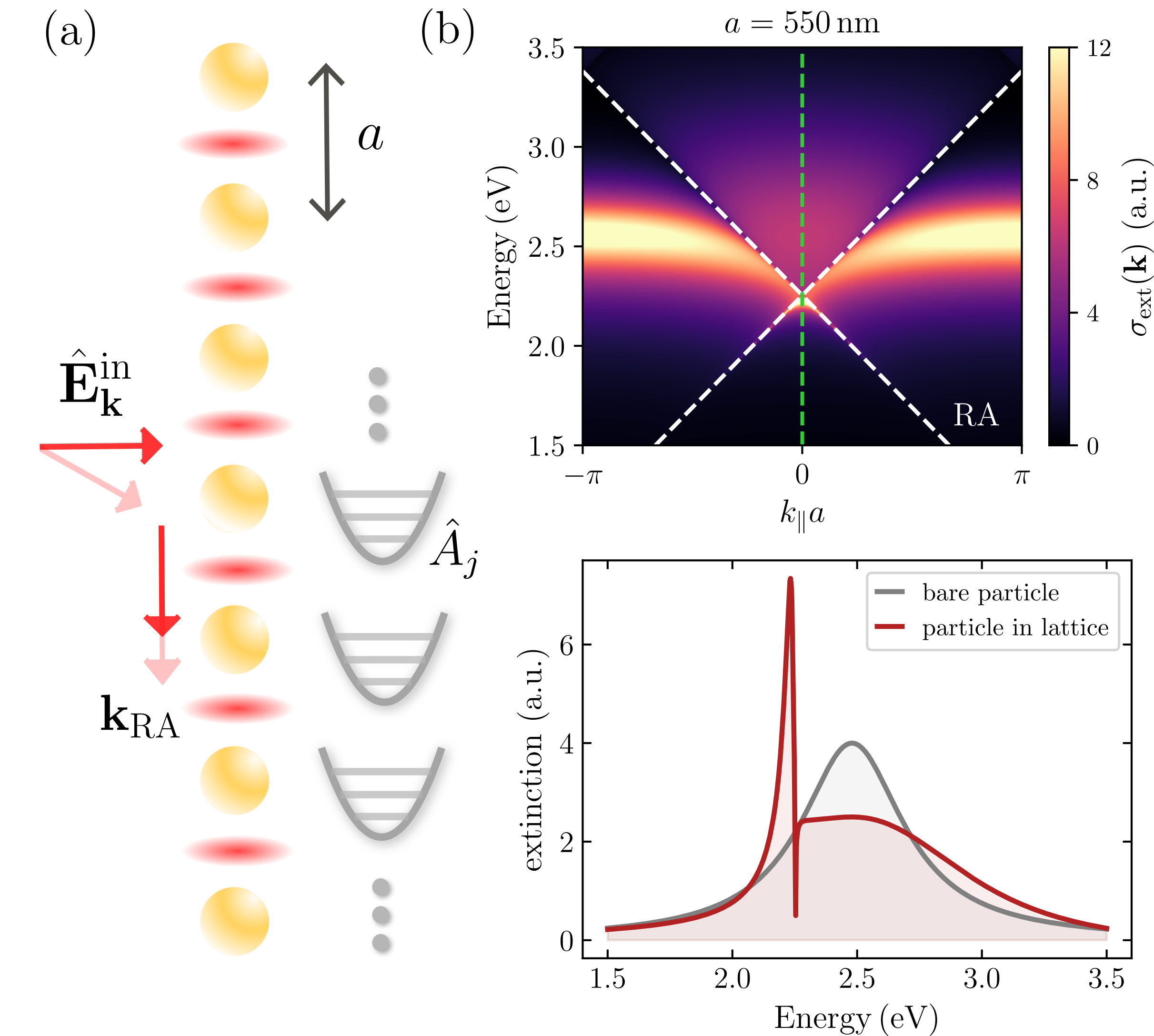}
		\caption{(a) 1D array of MNPs with lattice spacing $a$, modeled as quantum harmonic oscillators with annihilation operator $\hat A_j$, which are excited by an electric input field $\hat{\mathbf E}_\bk^\mathrm{in}(\br , t)$ incident at a wavevector $\bk$. In-plane diffractive modes at wavevectors $\mathbf{k}_\mathrm{RA}$ lead to the formation of SLRs (red shaded areas indicate regions of field enhancement). (b) Extinction spectrum for an array of $M=8\cdot 10^3$ MNPs with a resonance wavelength of $\lambda_0=500\:\mathrm{nm}$, a lattice spacing of $a=550\:\mathrm{nm}$, and a single-particle radiative linewidth of $\Gamma_0^\mathrm{rad}=0.5\:\mathrm{eV}$. The dashed white lines show the dispersion of the Rayleigh anomalies (RAs) corresponding to the first diffractive order ($m=1$). The lower panel shows a cross section through the extinction profile at $|\bk_\parallel|=0$ as indicated by the vertical green dashed line in the upper panel. The peak corresponds to the SLR while the dip arises from the RA. Here, the polarization of the incoming light field and the orientation of the dipoles are chosen orthogonal to the direction of the chain, i.e., $\mathbf{E}_0\parallel\boldsymbol\mu _0$, $\boldsymbol\mu_0 \perp \mathbf{r}_\Lambda$.}
	\label{fig1}
\end{figure}

Let us consider an array of $M$ MNPs located at positions $\br_j$ ($j\in \{ 1,\hdots, M\}$) which we model as quantum harmonic oscillators with resonance frequency $\omega_0$ and electric dipole operator $\hat{\bmu}_{0}^j=\bmu_0(\hat A_j^{\phantom\dagger}+\hat A_j^\dagger)$. Here, the annihilation and creation operators follow the standard bosonic commutation relation $[\hat A_j^{\phantom{\dagger}},\hat A_{j'}^\dagger]=\delta_{j,j'}$. We consider a regular lattice of particles spaced between each other by a distance $a$, as illustrated in Fig.~\ref{fig1}(a). While the theoretical treatment employed in the following holds for both 1D and 2D lattices, for the sake of simplicity we will focus on the 1D case. \revise{We note, however, that the physics of surface lattice resonances  can exhibit substantial distinctions  between 1D and 2D geometries due to differences in dipole coupling structure, mode multiplicity and convergence, as discussed for instance in Refs.~\cite{zou2004narrow, cherqui2019plasmonic}.} The lattice is assumed to be quasi-infinite which allows us to impose periodic boundary conditions. The array of nanoparticles is then described by the Hamiltonian
\begin{align}
\hat{\calH}_\text{MNP}=\hbar\sum_{j=1}^M \omega_0^{\phantom{\dagger}} \hat A_j^\dagger \hat A_j^{\phantom{\dagger}}.
\end{align}
The MNPs are immersed in the infinite set of electromagnetic free-space modes in a fictitious quantization box of mode volume $\mathcal V$, described by 
\begin{align}
\hat{\calH}_\text{vac}=\hbar\sum_{\bk, \lambda} \omega_\bk^{\phantom{\dagger}} \hat a_{\bk, \lambda}^\dagger \hat a_{\bk,\lambda}^{\phantom{\dagger}},
\end{align}
where $\hat a_{\bk,\lambda}^{\phantom{\dagger}}$ ($\hat a_{\bk,\lambda}^{\dagger}$) is the photon annihilation (creation) operator for a given electromagnetic field mode,  $\omega_\bk=c|\bk|$ is the vacuum photonic dispersion, and the sum goes over all the wavevectors $\bk$ which can be decomposed into in-plane and out-of-plane components $\bk=(\bk_\parallel, k_\perp)^\top$, as well as over the two orthogonal polarizations associated with each $\bk$ mode which are denoted by the index $\lambda$. The interaction between the MNPs and the bath of electromagnetic modes is described within the rotating wave approximation by
\begin{align}
\label{eq:hint}
\hat{\calH}_\text{int}=\hbar \sum_{j=1}^M\sum_{\bk,\lambda} \left(g_{\bk, \lambda}^{\phantom{\dagger}}  \hat A_j^\dagger \hat a_{\bk,\lambda}^{\phantom{\dagger}} \me^{\mi \bk\cdot\br_j}+\mathrm{H.c.}\right),
\end{align}
with the light-matter coupling strength $g_{\bk, \lambda}=-\mi \mathcal{E}_\bk (\beps_{\bk,\lambda}\cdot \bmu_0)/\hbar$, where the zero-point electric field amplitude is given by $\mathcal{E}_\bk=\sqrt{\hbar\omega_\bk/(2\epsilon_0 \mathcal{V})}$ ($\epsilon_0$ is the vacuum permittivity). 

Our objective is now to find a description for the response of the MNP array to a plane-wave input field $\hat{\mathbf E}_{\bk}^\mathrm{in}(\br, t)=-\mathrm{i}\mathbf E_0(\mathrm{e}^{-\mi\bk\cdot\br}\hat{a}_{\bk,\lambda}^\mathrm{in}(t)-\mathrm{H.c.})$, with polarization vector $\mathbf E_0$. This can be achieved by deriving the equations of motion for the dipole operators as well as for the the electric field amplitudes in the Heisenberg picture, followed by an elimination of the electromagnetic degrees of freedom, to obtain a reduced description for the matter part only [see Supplemental Material (SM) for detailed steps of the derivation]. This procedure is analogous to the derivation of generalized Langevin equations \cite{zwanzig2001nonequilibrium} and closely resembles the treatment of collective emission of atomic dipoles in quantum optics \cite{dicke1954coherence, lehmberg1970radiationi, Lehmberg1970radiationii, gardiner2010quantum}. Crucially, however, it does not invoke the Markovian assumption typically applied to such systems, thereby incorporating all retardation effects which are essential for obtaining the narrow lineshape associated with SLRs.
Given the harmonic nature of $\hat{\calH}_\text{MNP}$, $\hat{\calH}_\text{vac}$, and $\hat{\calH}_\text{int}$, we afford an analytical expression for the dipole amplitudes in Fourier domain
 \begin{align}
\label{eq:inputoutput}
\hat A_{\bq} (\omega) =\frac{\hat A_\bq^\mathrm{in}(\omega)}{\mi (\omega_0-\omega)+\Gamma_0^\mathrm{rad}/2-\mi\calS_\bq(\omega)},
\end{align}
where the index $\bq$ describes the quasi-momentum vector of the excitations along the array. Here, $\hat{A}_\bq^\mathrm{in}(\omega)$ represents the input operator, and $\Gamma_0^\mathrm{rad} = |\boldsymbol{\mu}_0|^2 \omega_0^3/(3 \pi c^3 \hbar \epsilon_0)$ is the radiative linewidth of an individual particle. The key quantity is the term $\calS_\bq(\omega)$, which involves a sum over all lattice vectors $\br_\Lambda$: 
\begin{align}
\calS_\bq(\omega) = \frac{|\bmu_0|^2 \omega^2}{c^2 \hbar \epsilon_0} \left( \sum_{\Lambda \setminus \{0\}}\me^{-\mi \bq \cdot \br_\Lambda}  \beps_{\boldsymbol \mu} \cdot \mathbf{G} (\br_\Lambda, \omega) \cdot \beps_{\boldsymbol \mu} \right), 
\end{align}
where $\Lambda$ describes the set of all lattice displacements from a central particle in the lattice and the summation does not include the zero displacement (see SM). This describes a discrete Fourier transform of the electromagnetic dyadic Green's tensor $\mathbf G (\br, \omega)$ with respect to all lattice vectors. Here, $\beps_{\boldsymbol \mu}=\bmu_0/|\bmu_0|$ denotes the unit vector in the direction of the nanoparticle dipole moment, thereby selecting the component of the Green's tensor mediating the interaction. Using that the dipole operator can be decomposed into positive and negative frequency components, $\hat{\boldsymbol \mu}_\bq\equiv\hat{\boldsymbol\mu}_\bq^{(-)}+\hat{\boldsymbol\mu}_\bq^{(+)}$, and leveraging the relation between the input operator and the negative frequency component of the electric field operator $\hat A_\bq^\mathrm{in} (\omega)=\mi \boldsymbol{\mu}_0\cdot \hat{\bE}_\bq^{\mathrm{in}, (-)}(\omega)$, we can express the MNP dipole operator in terms of the input electric field as follows: $\hat{\boldsymbol\mu}_\bq^{(-)}(\omega)=\label{eq:pol}
\boldsymbol{\alpha}_\bq^{\mathrm{eff}}(\omega)\hat{\bE}_\bq^{\mathrm{in}, (-)}(\omega)$, where the proportionality matrix
\begin{align}
\label{eq:pol}
\boldsymbol{\alpha}_\bq^{\mathrm{eff}}(\omega)=\frac{ \bmu_0 \cdot\bmu_0^\top }{(\omega_0-\omega)-\mi\Gamma_0^\mathrm{rad}/2-\calS_\bq (\omega)}
\end{align}
can be identified as the polarizability tensor of the array.
\begin{figure}[t]
	\centering
		\includegraphics[width=1\columnwidth]{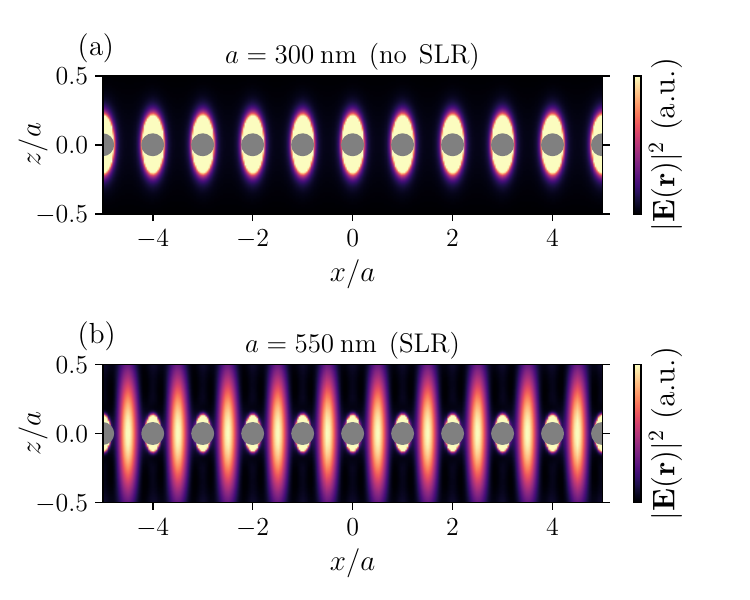}
		\caption{Intensity profile $|\mathbf E (\mathbf r)|^2$ radiated by a 1D chain of MNPs (indicated as silver dots) with a resonance wavelength of $\lambda_0=500\:\mathrm{nm}$ and a radiative linewidth of $\Gamma_0^\mathrm{rad}=0.5\:\mathrm{eV}$, for a lattice spacing of (a) $a=300\:\mathrm{nm}$, and (b) $a=550\:\mathrm{nm}$. The field distribution is plotted in the $xz$-plane and we have assumed normally-incident illumination of the array ($\theta_\mathrm{inc}=0$) with a polarization vector  along the $y$ direction, i.e., $\mathbf{E}_0 \parallel \boldsymbol\mu_0\parallel \mathbf{e}_y$. In (a), the electric field is evaluated at the frequency corresponding to the original LSP resonance ($\omega_0=2\pi c/\lambda_0$), while in (b) it is evaluated at the red-shifted SLR frequency.}
	\label{fig2}
\end{figure}

This is in full agreement with the result obtained from the classical electromagnetical description (see, e.g., Eq.~(10) of Ref.~\cite{cherqui2019plasmonic}). More generally, the effective polarizability tensor of the MNP lattice can always be related to the single-particle polarizability $\boldsymbol\alpha_0(\omega)$ via $[\boldsymbol\alpha_\bq^{\mathrm{eff}}]^{-1}(\omega)=\boldsymbol{\alpha}_0^{-1}(\omega)-\mathcal{S}_\bq(\omega)/(\bmu_0\cdot\bmu_0^\top )$. The extinction spectrum (describing the combined effects of absorption and scattering) of the nanoparticle array is defined as \cite{cherqui2019plasmonic}
\begin{align}
\sigma_\bk^\mathrm{ext} (\omega)= \frac{4\pi k}{|\mathbf{E}_0|^2}\,\mathrm{Im}\left[\mathbf{E}_0^{\top}\cdot \boldsymbol{\alpha}_{\bk_\parallel}^{\mathrm{eff}}(\omega) \cdot\mathbf{E}_0\right],
\end{align}
where the polarization of the incoming field $\mathbf{E}_0$ picks the components of the polarizability tensor that gets excited. The extinction spectrum of a 1D chain of MNPs with a lattice spacing of $a=550\,\mathrm{nm}$ is shown in the upper panel of Fig.~\ref{fig1}(b). A cross section through the extinction spectrum at zero quasimomentum $|\bk_\parallel| = 0$ is plotted below and compared with the extinction of a bare particle, revealing a substantial modification of the spectral response. SLRs are characterized by a dramatically narrowed peak with a much larger extinction as compared to the bare particle response, accompanied by a broader background at higher energies stemming from the bare original LSP response. In the following we provide some intuition on SLRs. SLRs can be interpreted as arising from the coupling of the LSPs with in-plane diffractive orders of the lattice given by the so-called \textit{Rayleigh anomalies} $\bk_\mathrm{RA}= \bk_\parallel \pm \boldsymbol{\mathcal{G}}$ where $\boldsymbol{\mathcal{G}}$ is a vector of the reciprocal lattice [in a single dimension one has $\mathcal{G}=m(2\pi/a)$ where $m=1,2,\hdots$] \cite{levan2019enhanced}. In particular, for an array probed by incident light with wavevector $\bk$ (with, in general, both in- and out-of-plane components), the emergence of SLRs requires that the wavevector of the diffracted light lies fully in the plane of the array, that is, with wavevector component perpendicular to the array plane vanishing; thus, conservation of energy of the incident light and the diffracted light implies that$\sqrt{k^2-|\mathbf{k}_\mathrm{RA}|^2}=0$.  This criterion implies that for light incident at an angle $\theta_\mathrm{inc}$ with $|\bk_\parallel |=k \sin\theta_\mathrm{inc}$, the condition to observe SLRs is given by $\lambda=(a/m)(1\pm\sin\theta_\mathrm{inc})$. The dispersion of the  first diffractive order  $m= 1$ is illustrated as the white dashed lines in the upper panel of Fig.~\ref{fig1}(b). The coupling of the diffractive order with the LSP leads to an avoided crossing with a Fano-type lineshape of the SLR.

The characteristic properties of SLRs can be further illustrated by plotting the intensity profile $|\mathbf E(\mathbf r)|^2$ radiated by the MNP array in response to the external illumination, as shown in Fig.~\ref{fig2} for normally incident light (corresponding to $\bk_\parallel=0$). For lattice spacings at which the SLR condition is not met [cf.~Fig.~\ref{fig2}(a)], the electric field  is dominated by the local plasmonic fields of the individual MNPs. On the other hand, fulfilling the SLR condition by choosing the appropriate lattice spacing leads to a dramatic change of the characteristics of the field distribution [cf.~Fig.~\ref{fig2}(b) for the first diffractive mode $m=1$]. The coupling to the diffractive modes allows a significant amount of the field to get `trapped' between the nanoparticles, forming a standing wave along the chain axis with antinodes of the field located between the particles. Higher-order diffractive modes would lead to multiple extrema of the field between the particles with a modulation of the field proportional to $\cos (2\pi m x/a)$.

 \section{Application 1: Molecular optomechanics with SLRs}
 \label{sec:molecularom}

 \begin{figure}[b]
	\centering
		\includegraphics[width=0.95\columnwidth]{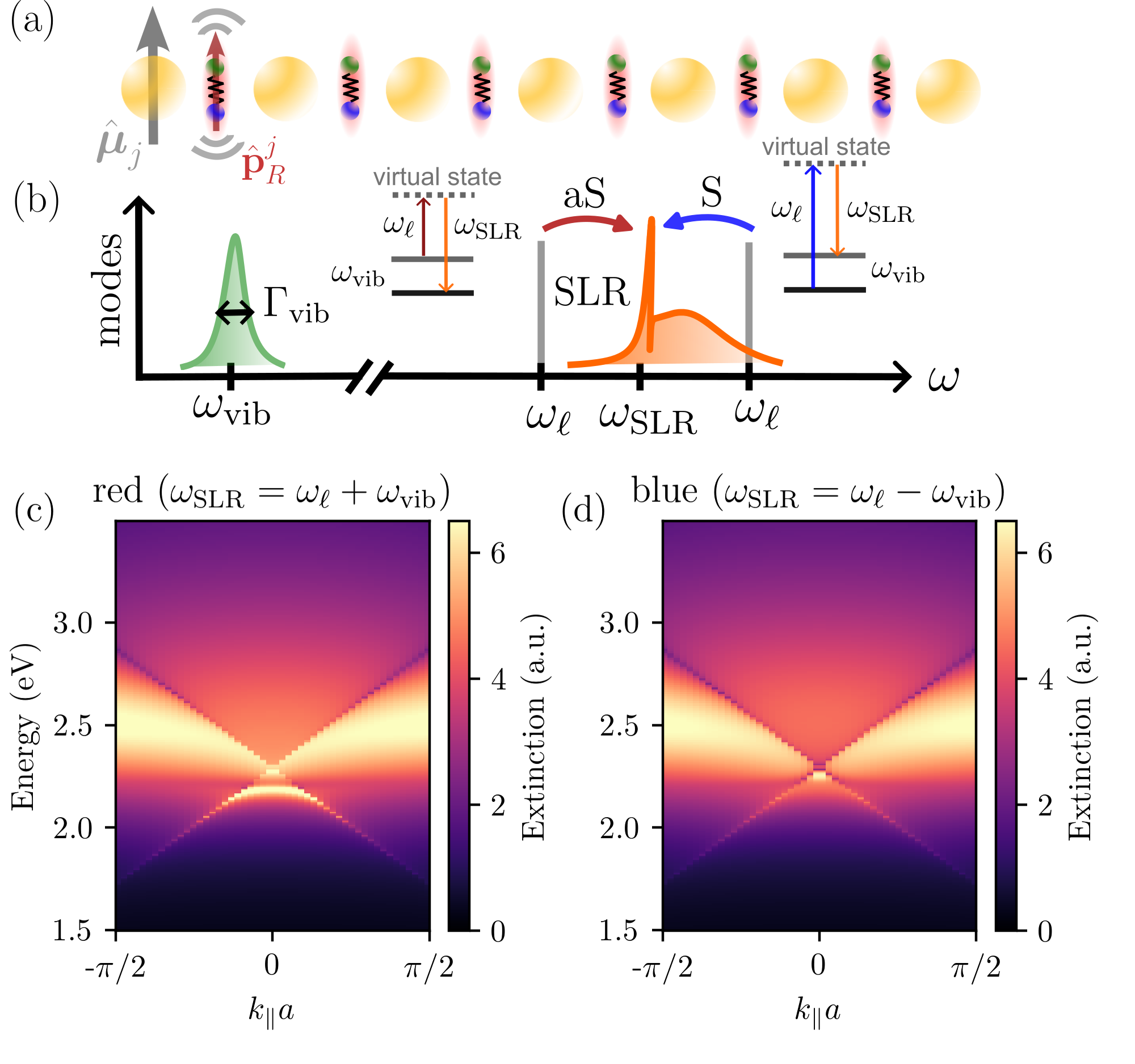}
		\caption{Molecular optomechanics with SLRs. (a) Sketch of array of molecular dipoles, illustrated as diatomic oscillators, characterized by their induced Raman dipoles $\hat{\mathbf{p}}_R^j$, interacting with the electric field produced by a set of MNPs, characterized by dipoles $\hat{\boldsymbol{\mu}}_j$. (b) Sketch of relevant interacting modes as well as processes leading to the enhancement of Stokes- and anti-Stokes scattering by the SLR mode. (c) and (d) show extinction spectra of the nanoparticle array in the red- and blue-detuned regime (regarding the laser frequency as compared to the SLR frequency at $k_\parallel = 0$). We assumed a nanoparticle linewidth of $\Gamma_0^\mathrm{rad}=0.5\,\mathrm{eV}$, a (collective) induced Raman dipole moment of $p=0.3\mu_0$, and have set $\Gamma_\mathrm{vib}=0$.}
	\label{fig3}
\end{figure}

The field of optomechanics is concerned with the interaction between light and mechanical motion. In a prototypical optomechanical setup, circulating photons exert a force on the mirrors due to radiation pressure \cite{aspelmeyer2014cavity}. The mutual interplay between light and motion gives rise to a variety of applications of optomechanical devices, e.g., as sensors \cite{krause2012high, fogliano2021ultrasensitive} or frequency-converters \cite{hill2012coherent}. Interestingly, an analogy can be found between the standard optomechanical setup and surface-enhanced Raman scattering (SERS) where the plasmonic field takes the role of the cavity mode and the molecular vibration takes that of the mechanical oscillator \cite{roelli2016molecular}. While SERS is typically understood as a classical process, parametric amplification phenomena experimentally observed in these systems sparked interest in recent years to translate concepts from quantum cavity optomechanics \cite{chen2021continuous}, leading to a quantum description of the interaction between molecular vibrations and plasmonic fields \cite{neuman2019quantum, patra2023molecular}. Given that molecular optomechanics often suffers from lossy plasmonic modes, it is worthwhile exploring whether the high $Q$-factors offered by SLRs can provide an advantage. In particular, this may help achieve the desired resolved sideband regime in optomechanics, which is characterized by vibrational frequencies exceeding the optical mode linewidth \cite{zou2005silver}, a crucial requirement for selectively addressing individual Stokes/anti-Stokes sidebands. \revise{SLRs have e.g.~been experimentally studied for SERS signal enhancement in Refs.~\cite{alam2020double, khinevich2023wavelength, andrzejewska2025towards} }. \revise{Although here, we focus on molecular vibrations, it is also noteworthy that optomechanical oscillations can occur directly within the nanoparticles themselves which has been observed via time-resolved spectroscopy and does not require the coupling to external emitters~\cite{hartland2011optical, juodenas2020effect}.}

To this end, we now consider an additional array of $M$ (identical) molecules at positions $\br_j^\mathrm{m}$ which are displaced by some fixed vector $\mathbf r_\mathrm{m}$ from the MNP lattice, i.e., $\br_j^\mathrm{m}=\br_j+\mathbf{r}_\mathrm{m}$. \revise{While such a perfectly ordered molecular array may be challenging to realize experimentally, we adopt this configuration here because it preserves translational symmetry and enables momentum conservation, thereby greatly simplifying the theoretical analysis. Further, we focus on
SLR modes with wave vectors much larger than the inverse intermolecular spacing, so that  we expect disorder-induced variations to effectively average out over these longer wavelengths. However, recent advances in nanoprinting techniques have demonstrated the ability to position organic molecules with high precision \cite{hail2019nanoprinting}, offering promising routes toward realizing such ordered hybrid architectures.} In the following we will aim at positioning the molecules into the  SLR `hotspots', as illustrated in Fig.~\ref{fig3}(a).
The general optomechanical Hamiltonian expresses as
\begin{align}
\hat{\calH}_\text{OM}=-\sum_j\hat{\mathbf  p}_j^R\cdot \hat\bE (\br_j^\mathrm{m}),
\end{align}
 with the induced Raman dipole of the molecule  $\hat{\mathbf p}_j^R=\alpha_\mathrm{m} (Q_j)\hat{\bE} (\br_j^\mathrm{m})$, where we consider the (scalar) molecular polarizability $\alpha_\mathrm{m} (\hat{Q}_j)$ to depend on a single quantized, harmonic nuclear coordinate $\hat{Q}_j=Q_\mathrm{zpm}(\hat b_j^\dagger+\hat b_j)$. Here, the zero-point motion is given by $Q_\mathrm{zpm}=\sqrt{\hbar/(2 m_\mathrm{vib}\omega_\mathrm{vib})}$ with $\omega_\mathrm{vib}$ the vibrational frequency and $m_\mathrm{vib}$ the reduced mass of the vibrational mode. The free Hamiltonian of the nuclear degrees of freedom is then simply given by $\hat{\mathcal{H}}_\mathrm{vib}=\sum_j\omega_\mathrm{vib} \hat{b}_j^\dagger \hat{b}_j^{\phantom{\dagger}}$. Taylor expansion of the polarizability with respect to the nuclear coordinates around the equilibrium configuration $Q_0$ then gives rise to a coupling between the electromagnetic field and the nuclear motion
 \begin{align}
 \alpha_\mathrm{m} (\hat{Q}_j)\approx \alpha_\mathrm{m} (Q_0)+  \left(\frac{\partial\alpha_\mathrm{m}(\hat{Q}_j)}{\partial Q_j}\right)_{Q_0}Q_\mathrm{zpm}(\hat{b}_j^\dagger+\hat{b}_j^{\phantom\dagger}). 
 \end{align}
 It is now convenient to treat the external illumination inducing the Raman dipoles as a classical field. This approximation corresponds
to substituting the field operator in the definition
of the induced Raman dipole operator $\mathbf p_j^R=\alpha_\mathrm{m}(Q_j)\bE (\br_j^\mathrm{m})$ by the classical local
field $\mathbf E$ at frequency $\omega_\ell$ \cite{zhang2021addressing}. This amounts to approximating
\begin{align}
\hat{\mathbf p}_j^R\approx \frac{1}{2}\left(\mathbf p_j^{\phantom{*}} \me^{-\mi\omega_\ell t}+\mathbf p_j^* \me^{\mi\omega_\ell t}\right)\left(\hat b_j^\dagger+\hat b_j^{\phantom\dagger}\right),
\end{align}
with the classical amplitudes of the induced Raman dipoles $\mathbf p_j = Q_\mathrm{zpm}(\partial\alpha_\mathrm{m}/\partial  Q_j)_{Q_0}\bE(\br_j^\mathrm{m})$. Under the assumption that all Raman dipoles are excited identically by the external laser field, i.e., $\mathbf p_j\equiv \mathbf p$, the optomechanical Hamiltonian describing the interaction of the induced Raman dipoles with the surrounding electromagnetic field can now be expressed in a linearized form as
\begin{align}
\label{eq:hom}
\hat{\mathcal{H}}_\mathrm{OM}= \hbar\sum_{j=1}^M\sum_{\bk,\lambda}\left[g_{\bk,\lambda}^{\mathrm{OM}}\hat{a}_{\bk,\lambda}\me^{\mi\bk\cdot\br_j^\mathrm{m}}\me^{\mi\omega_\ell t}+\mathrm{H.c.}\right](\hat b_j^\dagger+\hat b_j^{\phantom\dagger}),
\end{align}
with the optomechanical coupling strength $g_{\bk,\lambda}^\mathrm{OM}=-\mathcal{E}_\bk (\beps_{\bk,\lambda}\cdot\mathbf{p}^*)/(2\hbar)$. 

The Hamiltonian in Eq.~\eqref{eq:hom} describes resonant exchanges between molecular vibrations and the electromagnetic field, aided by the induced Raman dipoles which are oscillating at the laser-induced frequency $\omega_\ell$. Together with the original interaction Hamiltonian for the MNPs in Eq.~\eqref{eq:hint}, this describes now the coupling of two sets of dipoles to the electromagnetic field. Elimination of the electromagnetic bath will therefore give rise to an interaction between the MNPs and the molecular dipoles. 

We remark that this coupling can also arise from a collective molecular bright mode, which is formed when $N_\mathrm{m}$ molecules are positioned at the same location $\mathbf{r}_j$ or within a region much smaller than the wavelength of the electromagnetic field. The collective bright mode can be defined as
$\hat{B}_j^{\phantom{\dagger}} + \hat{B}_j^\dagger = \sum_{s=1}^{N_\mathrm{m}} \left(\hat{b}_{j,s}^{\phantom{\dagger}} + \hat{b}_{j,s}^\dagger \right)/\sqrt{N_\mathrm{m}}$,
where $\hat{b}_{j,s}$ and $\hat{b}_{j,s}^\dagger$ are the annihilation and creation operators for the $s$-th molecule's vibrational mode at position $\mathbf{r}_j$.

The formation of the collective bright mode leads to an enhancement of the optomechanical coupling strength at each position $\mathbf{r}_j$ by a factor of $\sqrt{N_\mathrm{m}}$. 
Proceeding analogously to the previous section by eliminating the electromagnetic field modes, one arrives at a set of coupled dipole equations in Fourier domain (see SM  for details) leading to a modified expression for the nanoparticle polarizability in the presence of the optomechanical coupling
\begin{align}
\label{eq:polopto}
\boldsymbol{\alpha}_\bq^{\mathrm{OM}}(\omega)=\frac{ \bmu_0 \cdot\bmu_0^\top }{(\omega_0-\omega)-\mi\Gamma_0^\mathrm{rad}/2-\calS_\bq^{\boldsymbol\mu} (\omega)-\Sigma_\bq^\mathrm{OM}(\omega)},
\end{align}
where $\calS_\bq^{\boldsymbol\mu} (\omega)$ refers to the lattice sum due to interactions among the MNPs, as already discussed in the previous section. Importantly, an additional term arises containing the optomechanical self-energy
\begin{align}
\label{eq:omlattice}
\Sigma_\bq^\mathrm{OM}(\omega)=\pm\frac{\mathrm{i}\mathcal{S}_\bq^{\mathrm{OM}}(\omega)^2}{\mathrm{i}(\omega_\ell\pm\omega_\mathrm{vib}-\omega)+(\Gamma_\mathrm{vib}\pm\gamma^\mathrm{rad}_{\mathrm{p},\pm})/2\mp\mi \calS_\bq^\mathbf{p} (\omega)},
\end{align}
with the phenomenological vibrational damping rate $\Gamma_\mathrm{vib}$. Here, we have introduced the ``optomechanical'' lattice sum describing interactions between Raman dipoles and MNPs
\begin{align}
 \calS_\bq^{\mathrm{OM}}(\omega) \! = \frac{\mu_{0} p\omega^2}{c^2\hbar\epsilon_0}\sum_{\Lambda}\me^{-\mi \bq\cdot (\br_\Lambda+\br_\mathrm{m})}\beps_{\boldsymbol{\mu}}\!\cdot\!\mathbf{G}(\br_\Lambda\!+\br_\mathrm{m},\omega)\!\cdot\! \beps_{\mathbf{p}}.
\end{align}
Note that, unlike the lattice sum of an individual array, which excludes the zero dispacement term to avoid divergence due to self-interaction, the optomechanical lattice sum incorporates the \( \br_\Lambda= 0 \) contribution, accounting  for the interaction between the Raman dipoles and the MNPs at the same lattice site.
In Eq.~\eqref{eq:omlattice}, the $\pm$ signs indicate whether the pump frequency is red-(blue-)detuned from the SLR frequency, thereby matching the anti-Stokes (Stokes) sideband, respectively (see SM for the derivation). The rates $\gamma_{\mathrm{p},\pm}^\mathrm{rad}=|\mathbf p |^2(\omega_\ell\pm\omega_\mathrm{vib})^3/(3\pi c^3\hbar\epsilon_0)$ describe the effects of the optomechanical interaction onto the molecular vibrational mode, leading to narrowing (heating) or broadening (cooling) due to the creation (annihilation) of vibrational quanta in the blue-(red-)detuned regimes, respectively. 
In addition, the optomechanical lattice sum also contains a term accounting for the interactions among the Raman dipoles in the molecular lattice
\begin{align}
\calS_\bq^{\mathbf{p}}(\omega)=\frac{|\mathbf{p}|^2\omega^2}{c^2\hbar\epsilon_0}\left(\sum_{\Lambda \setminus \{0\}} \me^{-\mi \bq \cdot\br_\Lambda} \beps_{\mathbf p}\cdot \mathbf G (\br_\Lambda,\omega) \cdot \beps_{\mathbf{p}}\right).
\end{align}
This term is in analogy with the collective interactions observed among Raman dipoles coupled to single plasmonic particles \cite{jakob2023giant}. While the above results are obtained under the rotating-wave approximation (RWA) between mechanical and photonic modes for both red- and blue-detuned driving, one should note an important subtlety in the blue-detuned regime. In the blue-detuned regime, i.e., $\omega_\ell > \omega_0$, the anti-Stokes sideband ($\omega_\ell + \omega_\mathrm{vib}$) can overlap with the asymmetric broad plasmonic background at higher energies [see Fig.~\ref{fig3}(b)], making the RWA not reliable anymore. In contrast, for red-detuned driving ($\omega_\ell < \omega_0$), the Stokes sideband ($\omega_\ell - \omega_\mathrm{vib}$) is well separated from the plasmonic background, and we find that optomechanical coupling effects in this regime can be robustly captured by the RWA. As a result, red detuning of the laser with respect to the SLR resonance should provide a clearer route to achieving the resolved sideband regime and optomechanical coupling in SLR-based platforms, making it the more promising scenario for experimental implementations.

It might be possible, however, that under specific design conditions—for example, if a suitable SLR is engineered such that its broad background resides at lower energies than $\omega_0$
 —the blue-detuned scenario becomes more favorable. \revise{For related photonic lattices, this can, e.g., be achieved by tuning the fill factor, which modifies the band structure and enables control over the relative position of leaky and bound-state modes~\cite{lee2019bandflip}.}

 \begin{figure}[b]
	\centering
		\includegraphics[width=1.0\columnwidth]{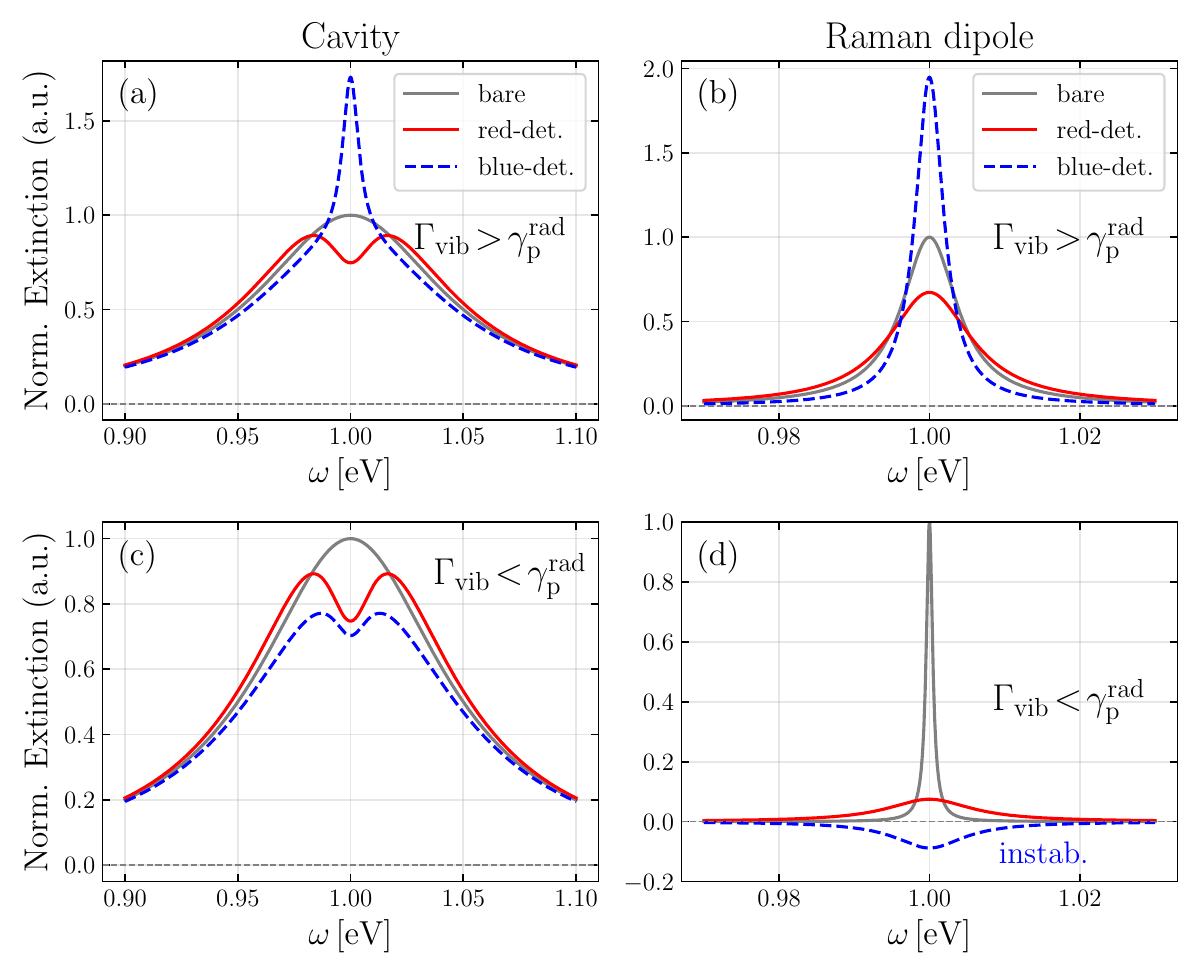}
		\caption{Comparison of cavity (left column) and Raman dipole extinction spectrum (right column) for (a), (b) $\Gamma_\mathrm{vib}>\gamma_\mathrm{p}^\mathrm{rad}$ and (c), (d) $\Gamma_\mathrm{vib}<\gamma_\mathrm{p}^\mathrm{rad}$ in the red- and blue-detuned regimes. The curves are normalized to the bare (grey) extinction spectra, obained in the absence of optomechanical interaction. We have set $\omega_0=\omega_\ell\pm\omega_\mathrm{vib}=1\,\mathrm{eV}$ and $\Gamma_0^\mathrm{rad}=0.1\,\mathrm{eV}$. In (a) and (b), we have set $\Gamma_\mathrm{vib}=10\gamma_\mathrm{p}^\mathrm{rad}=0.1\Gamma^0_\mathrm{rad}$, while in (c) and (d) these values are reversed.}
	\label{fig4}
\end{figure}

The resulting extinction spectra of the nanoparticle array are plotted in Figs.~\ref{fig3}(c) and (d) in the red- and blue-detuned regimes. Here, we placed the molecules in the regions of field enhancement between the MNPs, i.e., at $\br_\mathrm{m}=0.5 a\mathbf{e}_x$ and assumed dipole orientations $\mathbf p\!\parallel\!\bmu_0 \perp \!\br_\Lambda$. The red and blue sideband regime have distinct optical signatures due to the different underlying exchange processes.  In the anti-Stokes regime, where the laser is detuned below the SLR resonance (red-detuned) by the molecular vibrational frequency, the spectrum exhibits a dip in extinction  corresponding to optomechanically-induced transparency (OMIT), or in the strong coupling regime, a normal mode splitting due to coherent energy exchanges between the collective lattice resonance and the vibrational mode \cite{aspelmeyer2014cavity}.  In contrast, the blue sideband regime, with the pump laser detuned above the SLR resonance, shows a broadened peak in extinction and does not exhibit strong coupling.\\

To better understand the imprint of the optomechanical interaction onto the plasmonic mode and the Raman dipole  in the red- and blue-detuned regimes, we proceed by analyzing a simplified model that neglects the lattice sums in Eq.~\eqref{eq:polopto} and therefore only considers a single plasmonic mode coupled to a (collective) molecular Raman dipole.

Within this simplified description, one can distinguish two important regimes based on the magnitude of the laser-induced linewidth $\gamma_\mathrm{p}^\mathrm{rad}$ which are shown in Fig.~\ref{fig4}. In the first regime, where $\Gamma_\mathrm{vib}>\gamma_\mathrm{p}^\mathrm{rad}$ [see Figs.~\ref{fig4}(a), (b)], the red-detuned regime shows the characteristic OMIT while the blue-detuned regime shows an increase in extinction due to optomechanically-induced absorption (OMIA), arising from the amplification of the molecular vibrations \cite{aspelmeyer2014cavity}. The corresponding molecular extinction spectrum (vibrational mode dressed by laser field) in Fig.~\ref{fig4}(b) shows an increased (decreased) linewidth corresponding to cooling (heating) of the molecular vibrational mode, respectively. 

On the other hand, in the regime where $\Gamma_\mathrm{vib}<\gamma_\mathrm{p}^\mathrm{rad}$ [Figs.~\ref{fig4}(c), (d)], the molecular vibrational mode becomes unstable in the blue-sideband regime, a phenomenon known as parametric instability. This instability arises because the gain provided by the blue-detuned drive exceeds the intrinsic vibrational losses, causing a runaway amplification of the vibration. This  manifests itself as a negative extinction feature in the optical spectrum, indicating that the system is effectively amplifying, rather than attenuating, the probing field [see Fig.~\ref{fig4}(d)].

 \section{Application 2: Nonlinear switching of excitonic SLRs}
  \label{sec:excitonslr}
  
So far, the scope of this paper has been devoted to treating the nanoparticles providing the SLRs in the linear optical regime, \revise{corresponding to harmonic oscillators}. An interesting generalization of the framework is given by the extension to the nonlinear optical regime, which can, e.g., stem from anharmonicities of the particles. One possible realization of this consists in replacing the MNPs with a molecular material featuring tunable excitonic resonances such as proposed in Ref.~\cite{humphrey2016excitonic}. 

\begin{figure*}[t]
  \centering
  \includegraphics[width=1.0\textwidth]{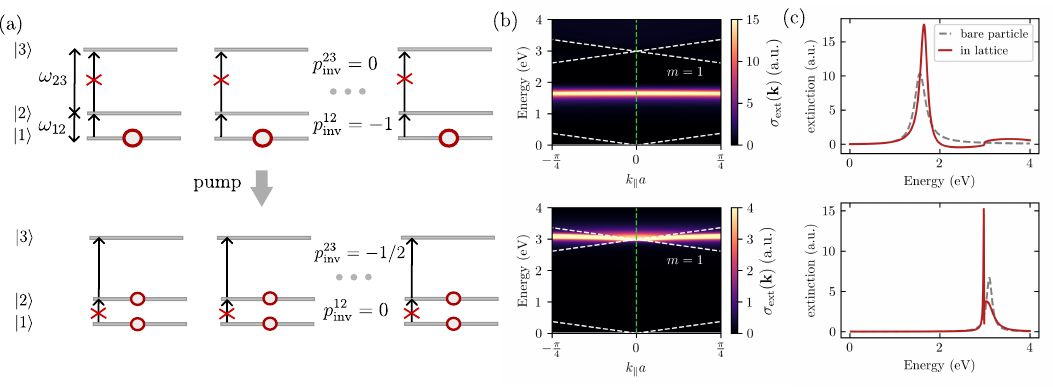}
		\caption{(a) Nonlinear SLR with array of saturable emitters: Pumping excitation into the $\ket{2}$ state shuts off the polarizability of the $\ket{1}\to\ket{2}$ transition while opening up the dipole transition from $\ket{2}\to\ket{3}$ whose frequency matches the first diffractive mode of the lattice. (b) Extinction spectrum of an array of $M=10^3$ three-level systems if all of the population is in the ground state (top) and if half of the population is pumped to the $\ket{2}$ state (bottom). The lattice spacing is $a=415\:\mathrm{nm}$. We chose the transition frequencies $\hbar\omega_{12}\approx 1.5\:\mathrm{eV},\, \hbar\omega_{13}\approx 3.0\:\mathrm{eV}$, as well as the radiative linewidths $\Gamma_{12}^\mathrm{rad}=\Gamma_{23}^{\mathrm{rad}}=0.25\:\mathrm{eV}$. The $\omega_{12}$ and $\omega_{23}$ transitions corresponds to a resonance wavelengths of $\lambda_{12}=800\:\mathrm{nm}$ and $\lambda_{23}=400\:\mathrm{nm}$, respectively. The dotted white lines show the frequencies corresponding to the Rayleigh anomalies. The vertical dashed green lines mark cross sections through the extinction spectrum at $k_\parallel =0$ which are shown in (c).}
	\label{fig5}
\end{figure*}

In the following, we will show that the nonlinearity of electronic transitions can be used to switch the SLR condition. This is enabled by `switching off' a transition that does not fulfill the Rayleigh anomaly condition and `switching on' a transition matching the Rayleigh criterion for transition between two excited states.  In particular, we consider a chain of anharmonic three-level chromophores described by states $\ket{\nu}_j$ where $j$ is the index describing the position of the particle in the lattice and $\nu$ describes the internal electronic label at energies $\hbar\omega_\nu$, $\nu=1,2,3$. These could for instance correspond the $S_0$, $S_1$ and $S_2$ states of an organic molecule. The transition dipole operator for a single molecule $j$ can then be expressed as $\hat{\boldsymbol\mu}^j=\sum_{\nu< \nu'}\hat{\boldsymbol\mu}_{\nu\nu'}^j$ where $\hat{\boldsymbol\mu}_{\nu\nu'}^j=\boldsymbol{\mu}_{\nu\nu'}(\hat\sigma_{j}^{\nu\nu'}+\mathrm{H.c.})$, with the operator for the $\nu\nu'$ transition $\hat\sigma_{j}^{\nu\nu'}=\ket{\nu}_j\bra{\nu'}_j$, and the dipole moment between two transitions is computed as $\boldsymbol{\mu}_{\nu\nu'}=\bra{\nu}_j \hat{\boldsymbol\mu}_{\nu\nu'}^j\ket{\nu'}_j$ which we assume to be identical for all molecules. We furthermore assume that all transition dipoles are oriented along the same direction. The free Hamiltonian of the excitonic lattice can then be expressed as
\begin{align}
\hat{\calH}_\text{ex}=\hbar \sum_{\nu=1}^3 \omega_\nu \sum_{j=1}^M  \ket{\nu}_j\bra{\nu}_j.
\end{align}
One can then derive an expression for the effective polarizability of the $\nu\nu'$ transition \revise{(which despite describing a signal that includes nonlinearities, can be computed in linear response)} (for details see SM) \cite{mukamel1995principles}
\begin{align}
\label{eq:polexc}
\boldsymbol{\alpha}_\bq^{\mathrm{eff}}(\omega)_{\nu\nu'}=-\frac{\mi p\ts{inv}^{\nu\nu'}(\boldsymbol{\mu}_{\nu\nu'}\cdot\boldsymbol{\mu}_{\nu\nu'}^\top)}{\left[\mi \mathcal{S}_\bq^{\nu\nu'} (\omega)p\ts{inv}^{\nu\nu'} -\mi(\omega-\omega_{\nu\nu'})+\Gamma_{\nu\nu'}^\mathrm{rad}/2 \right]},
\end{align}
where the label $\nu\nu'$ denotes the respective quantity for the $\nu\nu'$ transition (e.g., $\omega_{\nu\nu'}=\omega_{\nu'}-\omega_\nu$ is the frequency of the $\nu\nu'$ transition). The above expression is obtained under the assumption of equal population distribution in all emitters, i.e., $p_{\mathrm{inv},j}^{\nu\nu'}=\expval{\ket{\nu'}_j\bra{\nu'}_j-\ket{\nu}_j\bra{\nu}_j}=p_\mathrm{inv}^{\nu\nu'}$ and under the assumption of factorizability of populations and coherences, which, e.g., holds if populations evolve much slower than coherences and can be considered stationary. \revise{Here, the nonlinearity is implicit in the electric-field dependence of the populations.} Obviously, if $ p\ts{inv}^{\nu\nu'}=0$, the polarizability of a given transition is zero and will not contribute to the extinction spectrum. \revise{Furthermore, the validity of Eq.~\eqref{eq:polexc} is restricted to the regime $p\ts{inv}^{\nu\nu'}<0$, corresponding to the absence of population inversion.} The total polarizability is then simply obtained as a sum over the polarizabilites of all subtransitions $\boldsymbol{\alpha}_\bq^{\mathrm{eff}}(\omega)=\sum_{\nu<\nu'}\boldsymbol{\alpha}_\bq^{\mathrm{eff}}(\omega)_{\nu\nu'}$ \cite{fano1983pairs}. \revise{Let us emphasize that the expression in Eq.~\eqref{eq:polexc} assumes that the populations are fixed by their initial values and only (linear) coherences are evolving dynamically (see perturbative approach below and SM). At higher orders in the input field, populations will be driven dynamically and coherences and populations become coupled. In the section below, we show how nonlinear contributions in the input field can be incorporated into our formalism. Sophisticated treatments of nonlinear emitter dynamics have also been developed in the atomic array community~\cite{parmee2021bistable, scarlatella2024fate}.}

For illustrative purposes, let us consider a simplified scenario in the following where only the $1\to 2$ and $2\to 3$ transitions possess a non-vanishing dipole moment, i.e., $\mu_{13}=0$. The corresponding extinction spectra are plotted in Fig.~\ref{fig5}(b) while cross sections through the extinction spectrum at $|\bk_\parallel| = 0$ are shown in Fig.~\ref{fig5}(c). Here, we choose a scenario where the $\omega_{12}$ transition does not match the frequency corresponding to the Rayleigh criterion (around $|\bk_\parallel|=0$). The extinction spectrum  on top of Fig.~\ref{fig5}(c) shows a modification of the linewidth as well as the resonance frequency as opposed to a bare molecule due to the dipolar interactions in the lattice but there is no SLR. On the other hand, we chose the $\omega_{23}$ transition such that it matches the frequency of the $m=1$ Rayleigh anomaly modes at $|\bk_\parallel|=0$. The resulting extinction spectrum at $|\bk_\parallel| = 0$ now shows much stronger modification as opposed to the bare molecule spectrum, with a narrow SLR feature as discussed \revise{in the model section}. \\

\section{Pump-probe spectroscopy of nonlinear excitonic SLR dynamics}

\begin{figure}[t]
  \centering
  \includegraphics[width=0.95\columnwidth]{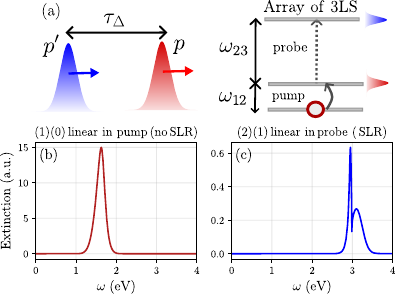}
		\caption{\revise{(a) Pump-probe spectroscopy of excitonic SLRs: Pump ($p$) and probe ($p'$) pulses excite the emitters at time delay $\tau_\Delta$ (at normal incidence $k_{\ell, \parallel}=0$). The central frequencies of the pump and probe envelopes in frequency domain are tuned to the $\omega_{12}$- and $\omega_{23}$-transitions, respectively. For the pulses, we choose a width of $40\,\mathrm{fs}$ in time domain and identical pulse envelopes $f_p(t)=f_{p'}(t)$. (b), (c) Linear extinction spectra in pump and probe field, at a time delay $\Delta\tau=50\,\mathrm{fs}$. The population created by the pump enables the $\omega_{23}$-transition, allowing the collective SLR resonance to emerge in the probe extinction spectrum. All other parameters are identical to Fig.~\ref{fig5}.}}
	\label{fig6}
\end{figure}

In this section, we show how dynamical evolution of the populations can be taken into account by performing a perturbative expansion in the input field \cite{mukamel1995principles, reitz2025nonlinear}. To this end, we model a pump–probe experiment with two impinging pulses, where a weak pump ($p$) prepares a nonequilibrium population distribution that is subsequently interrogated by a weak probe ($p'$) [see Fig.~\ref{fig6}(a)]. We assume the central frequency of the pump pulse to be (approximately) resonant with the $1\!\leftrightarrow\!2$ transition $\omega_p\approx \omega_{12}$, and the frequency of the probe pulse to be  (approximately) resonant with the $2\!\leftrightarrow\!3$ transition $\omega_{p'}\approx \omega_{23}$. Further, we assume the two transitions to be very off-resonant such that cross-couplings between them may be neglected. The (classical) input fields affecting the excitonic transitions of the $j$-th emitter can then be written as
\begin{subequations}
\begin{align}
\langle \hat{\sigma}_j^{12,\mathrm{in}}(t) \rangle &= \eta_p f_p(t) \, \me^{-\mathrm{i} \omega_p t}\me^{\mi\bk_\parallel\cdot\br_j}, \\
 \langle \hat{\sigma}^{23,\mathrm{in}}_j(t) \rangle &= \eta_{p'} f_{p'}(t-\tau_\Delta) \, \me^{-\mathrm{i} \omega_{p'} t}\me^{\mi\bk_\parallel\cdot\br_j},
\end{align}
\end{subequations}
where $\eta_{p,p'}$ describes the pulse amplitude, $f_{p,p'}(t)$ the temporal envelope (assumed Gaussian here), $\tau_\Delta$ is the delay time between the pulses, and we assume collinear incidence of pump and probe pulses at wavevector parallel to the array plane $\bk_\parallel$. Following Refs.~\cite{mukamel1995principles, reitz2025nonlinear}, we now expand all system operators $\hat{O}_\bq$ in terms of the pump and probe amplitudes (see Sec.~S5 of the SM for details)
\begin{align}
\langle\hat{O}_\bq\rangle=\sum_{n,m}\eta_p^n\eta_{p'}^m \langle\hat{O}_\bq\rangle^{(n)(m)}.
\end{align}
Assuming factorizability between coherences and populations, this yields a linearized set of equations up to $(n), (m)$-th order. In the following, we are interested in the linear response to the pump and the linear response to the probe, where the probe interrogates pump-induced second-order populations, yielding an overall third-order  correction. At first order, the pump can create coherence for the $1\!\leftrightarrow\!2$ transition: 
\begin{align}
\label{eq:firstorder}
\langle \hat{\s}_{\bq}^{12}(\omega)\rangle^{(1)(0)}=\frac{p_\mathrm{inv}^{12}(0) f_p(\omega-\omega_p)\delta_{\bq,\bk_{\parallel}}}{\mi \mathcal{S}_{\bq}^{12}(\omega) p\ts{inv}^{12}(0) -\mi(\omega-\omega_{12})+\Gamma_{12}^\mathrm{rad}/2 },
\end{align}
where the inversion variable for the $1\!\leftrightarrow\!2$ transition is fixed by the initial condition $p_\mathrm{inv}^{12}(0)$. This perturbative first–order expression is identical in form to the earlier result in Eq.~\eqref{eq:polexc}, which was obtained under the assumption of static populations. This correspondence arises, because in the perturbative framework, at first order, only coherences are created and therefore populations are fixed by their initial conditions. Population dynamics can only take place at second order in the pump where zero-momentum population is transferred into $\ket{2}$, rendering the \(2\!\leftrightarrow\!3\) inversion nonzero. This now leads to the third-order coherence linear in the probe field (see SM)
\begin{align}
\label{eq:thirdorder}
\langle \hat{\s}_{\bq}^{23}(\omega)\rangle^{(2)(1)}_{\Delta\tau}=\frac{[f_{p'}(\omega-\omega_{p'})_{\Delta \tau}\ast \langle \hat{\s}^{12,\dagger}\hat{\s}^{12}\rangle_{\bq-\bk_{\parallel}}^{(2)(0)}](\omega)}{\Gamma_{23}^\mathrm{rad}/2-\mi \mathcal{S}_{\bq}^{23}(\omega)  -\mi(\omega-\omega_{23})},
\end{align}
involving a convolution in frequency domain between the probe pulse and the second-order population created by the pump at zero quasi-momentum $\langle\hat{\s}^{12,\dagger}\hat{\s}^{12}\rangle_{\bq=0}^{(2)(0)}$ for $\bq=\bk_{\parallel}$. The above expression depends parametrically on the pulse delay time $\Delta\tau$ via the Fourier transform of the probe pulse.  This can be viewed as the third-order equivalent of Eq.~\eqref{eq:polexc}. Note however, that at third order, the only effect of the saturation is the renormalization of the input field via the created population in the numerator of Eq.~\eqref{eq:thirdorder}, while the lattice sum $\mathcal{S}_{\bq}^{23}(\omega)$ describing the interactions among the particles remains unmodified. In fact, the third–order term can be viewed as the next term in the expansion of the previous linear‐response expression, with the fixed inversion 
replaced by its pump–induced, dynamically-modified value, while modification of the interaction strength would show up as a higher-order effect. The resulting extinction spectra, both linear in pump and probe, are shown in Figs.~\ref{fig6}(b), (c). Importantly, we can see that the basic mechanism underlying SLR switching described in the previous section still remains valid even under the dynamic population assumption.

\section{Conclusions, outlook and extensions}
We have developed a quantum framework capable of describing dipolar arrays sustaining surface lattice resonances (SLRs) in the linear and nonlinear optical regime and with or without the presence of molecular emitters. We have highlighted applications of the formalism in molecular optomechanics as well as for excitonic SLRs where the nonlinearity of multi-level chromophores can be harnessed to switch the SLR condition by optical pumping. Importantly, we have bypassed \textit{ad hoc} few-mode approaches for the description of SLRs, which become cumbersome to handle in the presence of quantum emitters with complex multilevel structure.

In the first application, we discussed how high-$Q$ SLRs can \revise{provide} a pathway to bring molecular optomechanics in the resolved sideband regime and mediated the coupling between nanoparticle arrays and collective vibrational modes. 
In the second application, we demonstrated the versatility of SLRs in the nonlinear optical regime by considering arrays of multi-level emitters. Through optical pumping, the excitonic nonlinearity allowed for selective activation of specific SLR conditions corresponding to different electronic transitions. \revise{We have further shown that such switching can be observed and characterized in pump–probe experiments, for which we performed a perturbative expansion of the emitter dynamics up to third order, explicitly including population dynamics}. \revise{To the best of our knowledge, this constitutes the first description of pump–probe spectroscopy in SLRs that does not rely on phenomenological models, but instead derives directly from a microscopic treatment of the emitter–field interactions.}. This ability to dynamically control SLRs suggests potential uses in tunable photonic devices and optically driven switches, particularly in systems where reconfigurable optical properties are desirable.

For future endeavors, the quantized description of the electromagnetic field employed in our treatment could e.g., be used to investigate non-classical states of light such as \revise{squeezed} light input, with possible applications in quantum-enhanced sensing.  While so far, as a proof of principle, we have focused on extinction spectra, in principle we could also compute more complex material observables, like molecular populations through nonlinear optical spectroscopy, using mean-field approaches \cite{li2018mixed, fowler2022efficient, reitz2025nonlinear}.
Finally, while we have focused on a simplified treatment of the nanoparticles as harmonic oscillators in the dipole approximation, the approach can be extended to more complicated treatments, e.g., based on quasi-normal mode theory or higher-order multipoles \cite{miwa2021quantum}. \\

\section{Acknowledgments}

This research was primarily supported by the Center for Molecular Quantum Transduction (CMQT), an  Energy Frontier Research Center funded by the U.S.~Department of Energy, Office of Science, Office of Basic Energy Sciences, under Award no.~DE-SC0021314.

\bibliography{SLRRefs2}

\newpage
\onecolumngrid
\newcommand{\non}{\nonumber}

\newpage
\appendix

\begin{center}
  \textbf{\large Supplemental Material for ``Quantum theory of surface lattice resonances''}
\end{center}

\setcounter{table}{0}
\renewcommand{\thetable}{S\arabic{table}}
\setcounter{figure}{0}
\renewcommand{\thefigure}{S\arabic{figure}}
\setcounter{equation}{0}
\renewcommand{\theequation}{S\arabic{equation}}
\setcounter{section}{0}
\renewcommand{\thesection}{S\arabic{section}}

\section*{S1. Derivation of non-Markovian lattice response}

\label{sec:smlangevin}

The Heisenberg equations of motion for the electric field operator and dipole amplitudes are given by
\begin{subequations}
\begin{align}
\dot{\hat  a}_{\bk,\lambda} &= -\mi\omega_\bk {\hat  a}_{\bk,\lambda} -\mi g_{\bk,\lambda}^*\sum_j {\hat  A}_j\me^{-\mi \bk\cdot \br_j},\\
\dot {\hat  A}_j &=-\mi\omega_0 {\hat  A}_j - \mi\sum_{\bk,\lambda} g_{\bk,\lambda} {\hat  a}_{\bk,\lambda} \me^{\mi \bk\cdot \br_j}.
\end{align}
\end{subequations}
Formal integration of the electromagnetic field amplitudes
\begin{align}
 {\hat  a}_{\bk,\lambda} (t)={\hat  a}_{\bk,\lambda} (0)\me^{-\mi\omega_\bk t}-\mi g_{\bk, \lambda}^* \sum_{j=1}^M\me^{-\mi \bk\cdot \br_j}\int_0^t \td t'\, \me^{-\mi\omega_\bk (t-t')} {\hat  A}_j (t'),
\end{align}
and replacing them in the equation of motion for the dipole amplitudes yields
\begin{align}
\dot {\hat  A}_j=-\mi\omega_0 {\hat  A}_j -\mi\sum_{\bk,\lambda} g_{\bk, \lambda}\hat a_{\bk, \lambda} (0)\me^{-\mi\omega_\bk t}\me^{\mi \bk\cdot \br_j}-\sum_{\bk,\lambda} |g_{\bk, \lambda}|^2\sum_{j'}\me^{\mi \bk\cdot (\br_j-\br_{j'})}\int_0^t \td t' \me^{-\mi \omega_\bk (t-t')}{\hat  A}_{j'} (t').
\end{align}
The second term on the right-hand side in the equation above describes the input noise acting on the MNPs which we will denote in the following simply by ${\hat  A}_j^\mathrm{in}(t)= -\mi\sum_{\bk,\lambda} g_{\bk, \lambda}\hat a_{\bk,\lambda} (0)\me^{-\mi\omega_\bk t}\me^{\mi \bk\cdot \br_j}$. We define the Fourier transformation for the dipole operators (and likewise for all other quantities) as
\begin{align}
 {\hat  A}_j= \frac{1}{\sqrt M}\sum_\bq {\hat  A}_\bq \me^{\mi\bq\cdot \br_j},\qquad {\hat  A}_\bq=\frac{1}{\sqrt M}\sum_j {\hat  A}_j \me^{-\mi\bq\cdot \br_j},
\end{align}
such that 
\begin{align}
\label{eq:Aqelim}
\nonumber\dot {\hat  A}_\bq (t)&=-\mi\omega_0 {\hat  A}_\bq(t)+ {\hat  A}_\bq^\mathrm{in}(t)-\frac{1}{\sqrt{M}}\sum_{\bk,\lambda} | g_{\bk, \lambda}|^2\sum_{j,j'}\me^{-\mi\bq\cdot \br_j}\me^{\mi\bk\cdot(\br_j-\br_{j'})}\int_0^t \td t'\ \me^{-\mi\omega_\bk (t-t')}{\hat  A}_{j'} (t')\\
&=-\mi\omega_0 {\hat  A}_\bq(t)+ {\hat  A}_\bq^\mathrm{in}(t)-\sum_{\bq'}\underbrace{\left(\frac{1}{M}\sum_{\bk,\lambda} | g_{\bk, \lambda}|^2\sum_{j,j'}\me^{-\mi\bq\cdot \br_j}\me^{\mi\bk\cdot(\br_j-\br_{j'})}\me^{\mi\bq'\cdot\br_{j'}}\right)}_{\mathcal{K}^{\bq, \bq'}}\int_0^t \td t' \me^{-\mi\omega_\bk (t-t')}{\hat  A}_{\bq'} (t').
\end{align}
Let us consider only the expression in the bracket in the above equation which we denote by $\mathcal{K}^{\bq, \bq'}$
\begin{align}
\nonumber\mathcal{K}^{\bq, \bq'}&=\frac{1}{M}\sum_{\bk,\lambda} | g_{\bk, \lambda}|^2\sum_{j,j'}\me^{-\mi\bq\cdot \br_j}\me^{\mi\bk\cdot(\br_j-\br_{j'})}\me^{\mi\bq'\cdot\br_{j'}}=\\
&=\frac{1}{M}\sum_{\bk,\lambda}\frac{\omega_\bk}{2\hbar\epsilon_0 \mathcal{V}} | \bmu_0\cdot\beps_{\bk,\lambda}|^2 \sum_{j,j'}\me^{-\mi\bq\cdot \br_j}\me^{\mi\bk\cdot(\br_j-\br_{j'})}\me^{\mi\bq'\cdot\br_{j'}}.
\end{align}
Using the fact that $\bk\perp \beps_{\bk,1}\perp\beps_{\bk,2}$, the sum over the polarization degree of freedom $\lambda\in\{1,2\}$ can be carried out as
\begin{align}
\sum_{\lambda} | \bmu_0\cdot\beps_{\bk,\lambda}|^2=|\boldsymbol{\mu}_0|^2\left(1-\left(\beps_{\boldsymbol \mu}\cdot\beps_\bk\right)^2\right),
\end{align}
where we generally denote unit vectors as $\beps_{\mathbf v}=\mathbf v/ |\mathbf v |$ and $\beps_\bk$ is now the unit vector along the direction of $\bk$. In the limit of $\mathcal{V}\to \infty$, one can replace the discrete sum over $\bk$ vectors with an integral and move to spherical coordinates
\begin{align}
\frac{1}{\mathcal{V}}\sum_\bk\to\int\frac{\td^3 k }{(2\pi)^3}=\frac{1}{(2\pi c)^3}\int_0^\infty \td\omega_\bk\,\omega_\bk^2\int\td\Omega_\bk.
\end{align}
The sum can now be expressed as
\begin{align}
\mathcal{K}^{\bq, \bq'}&=\frac{\mu_0^2}{2(2\pi c)^3M\hbar\epsilon_0}\sum_{j,j'}\me^{-\mi\bq\cdot\br_j}\int_0^\infty\td\omega_\bk\,\omega_\bk^3\int\td\Omega_\bk\, [1-(\beps_{\boldsymbol \mu}\cdot\beps_\bk)^2]\me^{\mi\bk\cdot (\br_j-\br_{j'})}\me^{\mi\bq'\cdot\br_j}.
\end{align}
By making use of the identity $\beps_\bk\me^{\mi\bk\cdot\br}=\nabla\, \me^{\mi\bk\cdot\br}/(\mi k)$, we can substitute
\begin{align}
[1-(\beps_{\boldsymbol\mu}\cdot\beps_\bk)^2]=\left[1+\frac{\left(\beps_{\boldsymbol\mu}\cdot\nabla\right)^2}{k^2}\right],
\end{align}
and calculate the solid angle integral as 
\begin{align}
\left(1+\frac{(\beps_{\boldsymbol \mu}\cdot \nabla)^2}{k^2}\right)\int_0^\pi \td\theta_k\,\sin \theta_k\, \me^{\mi k|\br_j-\br_{j'}|\cos \theta_k }\int_0^{2\pi}\td\phi_k=4\pi \left[1+\frac{(\beps_{\boldsymbol \mu}\cdot \nabla)^2}{k^2}\right]\frac{\sin(k|\br_j-\br_{j'}|)}{k|\br_j-\br_{j'}|}.
\end{align}
This leads us to the expression for the sum 
\begin{align}
\nonumber\mathcal{K}^{\bq, \bq'}&=\frac{\mu_0^2}{(2\pi)^2c^3 \hbar\epsilon_0 }\frac{1}{M}\sum_{j,j'}\me^{-\mi\bq\cdot\br_j}\int_0^\infty \td \omega_\bk \,\omega_\bk^3 \left[1+\frac{(\beps_{\boldsymbol \mu}\cdot \nabla)^2}{k^2}\right]\frac{\sin k |\br_j-\br_{j'}|}{k |\br_j-\br_{j'}|}\me^{\mi\bq'\cdot\br_j}=\\
&=\int_0^\infty \td \omega_\bk\, \frac{\mu_0^2\omega_\bk^2}{\pi c^2\hbar \epsilon_0 }\frac{1}{M}\sum_{j,j'}\me^{-\mi\bq\cdot \br_j}\beps_{\boldsymbol\mu}\cdot \mathrm{Im}[\mathbf G (\br_j, \br_{j'},\omega_\bk)]\cdot \beps_{\boldsymbol\mu} \me^{\mi \bq' \cdot  \br_{j'}},
\end{align}
where we identified the imaginary part of the free space electromagnetic Green's tensor
\begin{align}
\mathbf{G} (\mathbf r_j, \mathbf{r}_{j'},\omega_\bk)\equiv\mathbf{G} (\mathbf r_j- \mathbf{r}_{j'},\omega_\bk)=\left(\identity+\frac{1}{k^2} \nabla\otimes\nabla\right)\frac{\me^{\iu k |\br_{j}-\br_{j'}|}}{4\pi|\br_{j}-\br_{j'}|},
\end{align}
which can be expressed in a more practical and explicit way as
\begin{align}
\mathbf{G} (\mathbf r ,\omega_\bk)=\frac{\me^{\mi k  r}}{4\pi k^2}\left[\left(\frac{k^2}{r}+\frac{\mi k }{r^2}-\frac{1}{r^3}\right)\identity+\left(-\frac{k^2}{r}-\frac{3\mi k}{r^2}+\frac{3}{r^3}\right)\frac{\br\otimes\br}{r^2}\right].
\end{align}
We can now express Eq.~\eqref{eq:Aqelim} as
\begin{align}
\dot {\hat  A}_\bq (t)&=-\mi\omega_0 {\hat  A}_\bq (t)+{\hat  A}_\bq^\mathrm{in} (t) -\sum_{\bq'}\int_0^\infty \td\omega_\bk\int_0^t \td t' \,\me^{-\mi\omega_\bk (t-t')}\frac{\mu_0^2\omega_\bk^2}{\pi c^2\hbar\epsilon_0}\frac{1}{M}\sum_{j,j'}\\\nonumber
&\times\me^{-\mi \bq\cdot\br_j}\beps_{\boldsymbol \mu}\cdot \mathrm{Im}[\mathbf G (\br_j,\br_{j'},\omega_\bk) ]\cdot \beps_{\boldsymbol\mu} \me^{\mi\bq'\cdot \br_{j'}}{\hat  A}_{\bq'}(t').
\end{align}
Due to the translational invariance of the lattice, double sums can be reduced to single sums
\begin{align}
\frac{1}{M}\sum_{j,j'}\me^{-\mi \bq\cdot\br_j}\beps_{\boldsymbol \mu}\cdot \mathrm{Im}[\mathbf G (\br_j,\br_{j'},\omega_\bk) ]\cdot \beps_{\boldsymbol\mu} \me^{\mi\bq'\cdot \br_{j'}}&=\frac{1}{M}\sum_{j,j'}\me^{-\mi\bq\cdot(\br_j-\br_{j'})}\mathrm{Im}\left[\mathbf G (\br_j,\br_{j'},\omega_\bk)\right] \me^{\mi (\bq'-\bq)\cdot\br_{j'}}\\\nonumber
&=\sum_{\Lambda } \me^{-\mi \bq\cdot\br_\Lambda}\beps_{\boldsymbol \mu}\cdot \mathrm{Im}[\mathbf G (\br_\Lambda,\omega_\bk) ]\cdot \beps_{\boldsymbol\mu}\,\delta_{\bq, \bq'},
\end{align}
where the sum now only depends on all relative distances with respect to a central point in the lattice which we denote by the sum over the set of lattice displacements $\Lambda$. With this, we obtain the equation of motion
\begin{align}
\label{eq:fourier}
\dot {\hat  A}_\bq (t)&=-\mi\omega_0 {\hat  A}_\bq (t)+{\hat  A}_\bq^\mathrm{in} (t) -C\int_0^\infty \td\omega_\bk\, \omega_\bk^2\int_0^t \td t' \me^{-\mi\omega_\bk (t-t')}\sum_{\Lambda} \me^{-\mi\bq\cdot \br_\Lambda}\beps_{\boldsymbol \mu}\cdot \mathrm{Im}[\mathbf G (\br_\Lambda,\omega_\bk) ]\cdot \beps_{\boldsymbol\mu} {\hat  A}_{\bq}(t'),
\end{align}
where we denoted the constant prefactors by $C=\mu_0^2/(\pi c^2\hbar\epsilon_0 )$. In Fourier space, the equation of motion for a single frequency component $\omega$ expresses as
\begin{align}
\nonumber -\mi\omega {\hat  A}_\bq (\omega)=-\mi\omega_0 {\hat  A}_\bq (\omega) + {\hat  A}_\bq^\mathrm{in}(\omega)&-C\int_0^\infty\td\omega_\bk \,\omega_\bk^2 \int_0^t \td t' \me^{-\mi \omega_\bk (t-t')}\\&\times\sum_{\Lambda} \me^{-\mi \bq\cdot \br_\Lambda} \beps_{\boldsymbol{\mu}}\cdot \mathrm{Im}\left[\mathbf G(\br_\Lambda,\omega_\bk)\right]\cdot \beps_{\boldsymbol{\mu}}
{\hat  A}_\bq (\omega)\,\me^{\mi\omega(t-t')}.
\end{align}
For $t\to\infty$, making use of the Sokhotski–Plemelj theorem
\begin{align}
\lim_{\epsilon\to 0^+}\int \td \omega_\bk \int_0^\infty \td s\, \me^{- \mi (\omega_\bk-\omega_0-\mi\epsilon)s}=\int \td \omega_\bk\left[\pi \delta (\omega_\bk-\omega_0)- \mi \mathcal{P} \left(\frac{1}{\omega_\bk-\omega_0}\right)\right],
\end{align}
where  $\mathcal P$ denotes the Cauchy principal value, and additionally making use of the (approximate) Kramers-Kronig relation for the dyadic Green's function \cite{buhmann2004vanderwaals, buhmann2007dispersion, dzsotjan2011dipole}
\begin{align}
\mathcal{P}\int_{-\infty}^\infty\td\omega_\bk\, \omega_\bk^2\, \frac{\mathrm{Im}[\mathbf G (\br_\Lambda,\omega_\bk)]}{\omega_\bk-\omega}\approx\pi\omega^2 \mathrm{Re}[\mathbf G (\br_\Lambda,\omega)],
\end{align}
one finally arrives at a simple expression relating the dipole amplitudes to the input term in Fourier space
\begin{align}
 {\hat  A}_{\bq} (\omega) &= \left[\mi (\omega_0-\omega)-\mi\frac{\mu_0^2\omega^2}{ c^2\hbar\epsilon_0}\left(\sum_{\Lambda} \me^{-\mi \bq \cdot\br_\Lambda} \beps_{\boldsymbol \mu}\cdot \mathbf G (\br_\Lambda,\omega) \cdot \beps_{\boldsymbol\mu}\right)\right]^{-1}  {\hat  A}_\bq^\mathrm{in}(\omega)\\\nonumber
&\equiv\left[\mi (\omega_0-\omega)-\mi\bar{\calS}_\bq(\omega)\right]^{-1} {\hat  A}_\bq^\mathrm{in}(\omega),
\end{align}
where the lattice sum 
\begin{align}
 \bar{\calS}_\bq(\omega)=\frac{\mu_0^2\omega^2}{ c^2\hbar\epsilon_0}\left(\sum_{\Lambda} \me^{-\mi \bq \cdot\br_\Lambda} \beps_{\boldsymbol \mu}\cdot \mathbf G (\br_\Lambda,\omega) \cdot \beps_{\boldsymbol\mu}\right)
 \end{align}
is the key quantity that one has to estimate as it describes the modification of the response due to the collective interactions among the particles in the lattice.
 
The imaginary part of the $\br_\Lambda=0$ contribution in the sum above is well defined and gives rise to the single-particle radiative decay rate \cite{novotnyhecht2006}, while we impose the diverging real part (corresponding to the dipole self-energy of an individual particle) to vanish
\begin{align}
\label{eq:damping}
\iu\beps_{\boldsymbol \mu}\cdot\mathrm{Im}\left[ \mathbf G (\br_\Lambda=0,\omega)\right] \cdot \beps_{\boldsymbol\mu}=\frac{\iu\omega}{6\pi c}, \qquad \mathrm{Re}\left[\mathbf G (\br_\Lambda=0, \omega)\right]\equiv 0,
 \end{align}
such that
 \begin{align}
\label{eq:inputoutput}
{\hat  A}_{\bq} (\omega) =\left[\mi (\omega_0-\omega)+\Gamma_0^\mathrm{rad}/2-\mi\calS_\bq(\omega)\right]^{-1} {\hat  A}_\bq^\mathrm{in}(\omega),
\end{align}
in terms of the single-particle radiative rate $\Gamma_0^\mathrm{rad}=(\mu_0^2 \omega_0^3)/(3 \pi c^3  \hbar  \epsilon_0)$. Here, the reduced lattice sum $\calS_\bq(\omega)$ now excludes the zero-displacement contribution and we made a Markovian assumption for the single-particle radiative rate only by evaluating it around $\omega_0\approx\omega$. The condition that needs to be fulfilled for SLRs is that both the real and the imaginary part of the expression in the brackets becomes (close to) zero. For numerical calculations, it is useful to reexpress the constants appearing in front of the lattice sum in terms of the single-particle radiative decay rate as
\begin{align}
\calS_\bq (\omega) =3\pi \Gamma_0^{\mathrm{rad}} c \,\frac{\omega^2}{\omega_0^3}\left(\sum_{\Lambda \setminus \{0\}} \me^{-\mi \bq \cdot\br_\Lambda } \beps_{\boldsymbol \mu}\cdot \mathbf G (\br_\Lambda ,\omega) \cdot \beps_{\boldsymbol\mu}\right)=\frac{3}{2}\Gamma_0^{\mathrm{rad}}  \,\frac{\lambda_0^3}{\lambda(\omega)^2}\left(\sum_{\Lambda \setminus \{0\}} \me^{-\mi \bq \cdot\br_\Lambda } \beps_{\boldsymbol \mu}\cdot \mathbf G (\br_\Lambda ,\omega) \cdot \beps_{\boldsymbol\mu}\right),
\end{align}
with $\lambda(\omega)=2\pi c/\omega$. \revise{Equivalently, we remark that one could carry out the derivation entirely in real space and apply the lattice Fourier transform only at the end.}

 \section*{S2. Coupling to a dipolar array of emitters}

\begin{figure*}[t]
  \centering
  \includegraphics[width=0.95\textwidth]{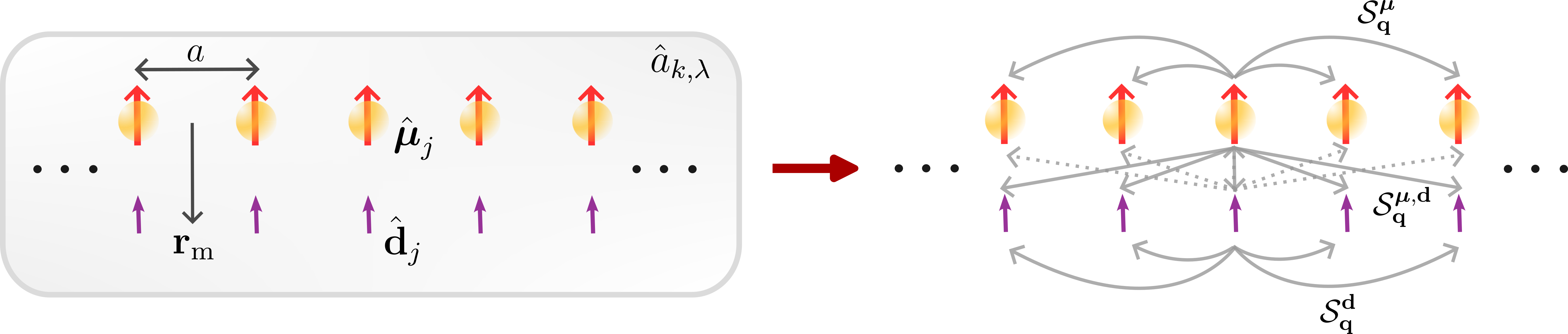}
		\caption{ \textbf{Schematics of elimination procedure.} Illustration of a nanoparticle array ($\hat{\bmu}_j$) and an array of dipolar emitters ($\hat{\mathbf{d}}_j$) which are commonly coupled to the background electromagnetic field modes $\hat{a}_{\mathbf{k}, \lambda}$. Elimination of the electromagnetic field leads to effective interactions both within the arrays and between the nanoparticle and emitter array, which are encompassed in the corresponding lattice sums. }
	\label{figS1}
\end{figure*}

Let us now consider the additional coupling of the nanoparticle array to an (e.g., molecular) array of dipolar emitters. We use this as an instructive case to facilitate the derivation of the optomechanical interaction in the following section. We assume the individual emitters with resonance frequency $\omega_d$ located at positions $\br_j^\mathrm{m}=\br_j+\br_\mathrm{m}$ described by the dipole operator $\hat{\mathbf{d}}_j=\mathbf{d}_0\left(\hat{B}_j+\hat{B}_j^\dagger\right)$ with dipole moment $\mathbf{d}_0$ and annihilation/creation operators $[\hat{B}_j^{\phantom{\dagger}},\hat{B}_{j'}^\dagger]=\delta_{j,j'}$ and typically $|\mathbf d_0|\ll|\boldsymbol\mu_0|$. We also assume that the emitter array has the same periodicity as the nanoparticle array. We start again by writing equations of motion for dipole operators as well as the electromagnetic field operators
\begin{subequations}
\begin{align}
\dot{\hat a}_{\bk,\lambda} &= -\mi\omega_\bk \hat{a}_{\bk,\lambda} -\mi g_{\bk, \lambda}^{\bmu,*}\sum_j \hat{A}_j\me^{-\mi\bk\cdot\br_j}-\mi g_{\bk,\lambda}^{\mathbf{d},*}\sum_j \hat{B}_j\me^{-\mi\bk\cdot\br_j^\mathrm{m}},\\
\dot{\hat{A}}_{j} &=-\mi\omega_0\hat{A}_j-\mi\sum_{\bk,\lambda}g_{\bk,\lambda}^{\bmu}\hat{a}_{\bk,\lambda}\me^{\mi\bk\cdot\br_j},\\
\dot{\hat{B}}_{j} &= -\mi\omega_d \hat{B}_j-\mi\sum_{\bk,\lambda}g_{\bk,\lambda}^{\mathbf{d}}\hat{a}_{\bk,\lambda}\me^{\mi\bk\cdot\br_j^m},
\end{align}
\end{subequations}
where we denoted the couplings by $g_{\bk, \lambda}^{\bmu}=-\mi \mathcal{E}_\bk (\beps_{\bk,\lambda}\cdot \bmu_0)/\hbar$ and $g_{\bk, \lambda}^{\mathbf{d}}=-\mi \mathcal{E}_\bk (\beps_{\bk,\lambda}\cdot \mathbf{d}_0)/\hbar$.
Again, we start by integrating out the electric field operators
\begin{align}
 {\hat  a}_{\bk,\lambda} (t)={\hat  a}_{\bk,\lambda} (0)\me^{-\mi\omega_\bk t}-\mi g_{\bk, \lambda}^{\bmu,*}\sum_{j=1}^M\me^{-\mi \bk\cdot \br_j}\int_0^t \td t'\, \me^{-\mi\omega_\bk (t-t')} {\hat  A}_j (t')-\mi g_{\bk, \lambda}^{\mathbf{d},*}\sum_{j=1}^M\me^{-\mi \bk\cdot \br_j^\mathrm{m}}\int_0^t \td t'\, \me^{-\mi\omega_\bk (t-t')} {\hat  B}_j (t'),
\end{align}
and replacing them in the equations for the dipole operators. The procedure is now largely identical to Sec.~S1 and consists in going to Fourier space both in the time and spatial domains. Finally, one ends up with an algebraic set of equations for the dipole operators
\begin{subequations}
\label{eq:dipolefreq}
\begin{align}
-\mi\omega \hat{A}_\bq (\omega) &=-(\mi\omega_0+\Gamma_0^\mathrm{rad}/2) \hat{A}_\bq (\omega)+\mi \calS_\bq^{\bmu}(\omega)\hat{A}_\bq (\omega)+\mi\calS_\bq^{\bmu,\mathbf{d}}(\omega)\hat{B}_\bq (\omega)+ \hat{A}_\bq^{\mathrm{in}}(\omega),\\
-\mi\omega \hat{B}_\bq (\omega)&=-(\mi\omega_d+\gamma_0^\mathrm{rad}/2) \hat{B}_\bq (\omega)+\mi \calS_\bq^{\mathbf{d}}(\omega)\hat{B}_\bq (\omega)+\mi\calS_\bq^{\mathbf{d},\bmu}(\omega)\hat{A}_\bq (\omega)+ \hat{B}_\bq^{\mathrm{in}}(\omega),
\end{align}
\end{subequations}
with the single-particle radiative rate of the emitters $\gamma_0^\mathrm{rad}=(|\mathbf d_0|^2\omega_d^3)/(3\pi c^3\hbar\epsilon_0)$, the lattice sums for nanoparticle and emitter lattices
\begin{align}
 \calS_\bq^{\bmu}(\omega) =\frac{|\bmu_0|^2\omega^2}{ c^2\hbar\epsilon_0}\sum_{\Lambda \setminus \{0\} }\me^{-\mi \bq\cdot \br_\Lambda}\beps_{\boldsymbol{\mu}}\cdot \mathbf{G}(\br_\Lambda,\omega)\cdot \beps_{\boldsymbol{\mu}}, \qquad  \calS_\bq^{\mathbf{d}}(\omega) =\frac{|\mathbf{d}_0|^2\omega^2}{ c^2\hbar\epsilon_0}\sum_{\Lambda \setminus \{0\} }\me^{-\mi \bq\cdot \br_\Lambda}\beps_{\mathbf{d}}\cdot \mathbf{G}(\br_\Lambda,\omega)\cdot \beps_{\mathbf{d}}.
\end{align}
as well as, crucially, the interaction between the nanoparticle and molecular lattices (assuming the dipole moments to be real)
\begin{align}
 \calS_\bq^{\bmu, \mathbf{d}}(\omega) = \frac{\mu_{0} d_0\omega^2}{c^2\hbar\epsilon_0}\sum_{\Lambda}\me^{-\mi \bq\cdot (\br_\Lambda+\br_\mathrm{m})}\beps_{\boldsymbol{\mu}}\cdot \mathbf{G}(\br_\Lambda+\br_\mathrm{m},\omega)\cdot \beps_{\mathbf{d}},
\end{align}
describing a sum over all interactions between nanoparticle and emitter array (see Fig.~\ref{figS1} for pictorial representation of lattice sums). The input noise terms $\hat{A}_\bq^\mathrm{in}(\omega), \hat{B}_\bq^\mathrm{in}(\omega)$ are defined analogously to Sec.~S1. Assuming only the nanoparticles to be driven by a non-zero average input electric field, i.e., $\expval{\hat{B}_\mathrm{in}(\omega)}=0$, the dipole amplitude of the nanoparticle array can be expressed as
\begin{align}
\hat{A}_\bq (\omega)=\frac{\mi\bmu_0\cdot \hat{\mathbf{E}}_\bq^{\mathrm{in}, (-)}(\omega)}{\mi(\omega_0-\omega)+\Gamma_0^\mathrm{rad}/2-\mi\calS_\bq^{\bmu} (\omega)+\frac{\calS_\bq^{\bmu,\mathbf{d}}(\omega)^2}{\mi(\omega_d-\omega)+\gamma_0^\mathrm{rad}/2-\mi\calS_\bq^\mathbf{d}(\omega)}}.
\end{align}

\section*{S3. Extinciton spectrum for different lattice constants}

\begin{figure*}[h]
  \centering
  \includegraphics[width=0.95\textwidth]{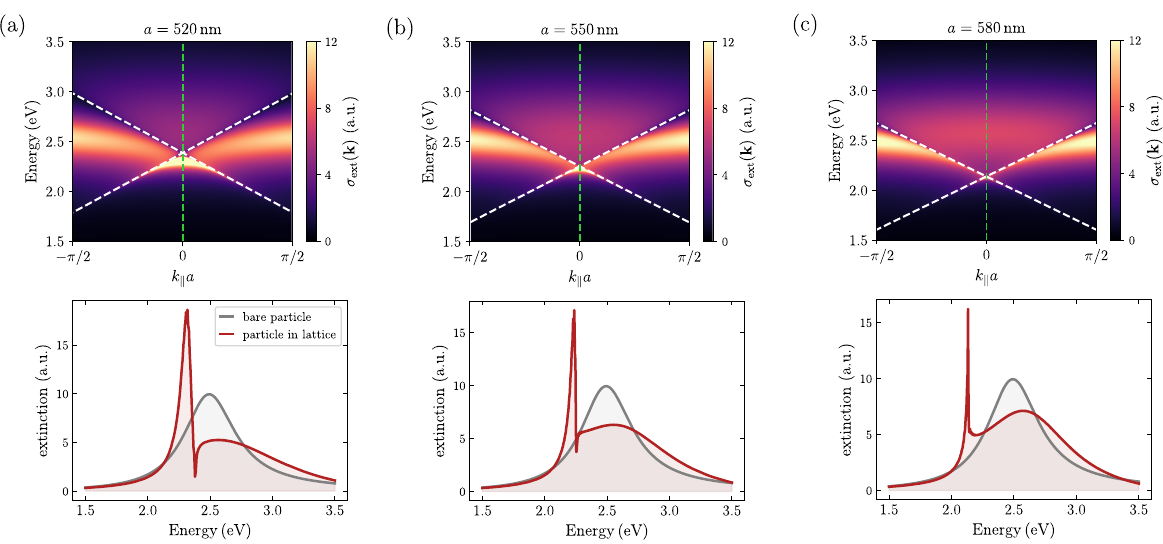}
		\caption{ \textbf{Extinction spectrum for different lattice constants.} Extinction spectra $\sigma_\mathrm{ext} (\bk)$ (top) as well as cross sections (illustrated by the green dashed curves) at normal incidence (bottom) for lattice constants (a) $a=520\,\mathrm{nm}$, (b) $a=550\,\mathrm{nm}$, and (c) $a=580\,\mathrm{nm}$ for a resonance wavelength of $\lambda_0=500\,\mathrm{nm}$ and radiative linewidth $\Gamma_0=0.5\,\mathrm{eV}$. As the lattice constant increases, the Rayleigh anomaly shifts to longer wavelength relative to the particle resonance, resulting in a progressive narrowing of the SLR linewidth.}
	\label{figq}
\end{figure*}

 \section*{S4. Optomechanical interaction}
\label{sec:supps3}

In the case of molecular optomechanics, the dipoles discussed in the previous section now correspond to Raman dipoles

\begin{align}
\label{eq:hamomsupp}
\hat{\Hm}_\mathrm{OM}=-\sum_j \hat{\mathbf p}_j^R\cdot \hat{\bE}(\br_j^\text{m}).
\end{align}
The Raman dipoles are induced by the electromagnetic field acting on the molecule $\hat{\mathbf{p}}_j=\alpha_\text{m}^j\hat{\mathbf{E}}(\br_j^\text{m})$ where the proportionality is given by the molecular polarizability $\alpha_\text{m}^j$. We assume the (scalar) polarizability to depend on a single normal-mode nuclear coordinate for each molecule $\alpha_\text{m}^j=\alpha_\text{m} (\hat{Q}_j)$ with $\hat{Q}_j=Q_\mathrm{zpm}(\hat b_j^\dagger+\hat b_j)$, where the zero-point motion is given by $Q_\mathrm{zpm}=\sqrt{\hbar/(2 m_\mathrm{vib}\omega_\mathrm{vib})}$ with $\omega_\mathrm{vib}$ and $m_\mathrm{vib}$ the vibrational frequency and reduced mass of the vibrational mode, respectively.

Due to the dependence of the Raman dipoles on the electric field, the Hamiltonian Eq.~\eqref{eq:hamomsupp} becomes quadratic in the electric field, making the treatment cumbersome. The Hamiltonian can however be linearized by the local field approximation (see, e.g., Ref.~\cite{zhang2021addressing} where one treats the external illumination inducing the Raman dipole as a classical field at laser frequency $\omega_\ell$
\begin{align}
\hat{\mathbf p}_j^R\approx \left(\mathbf p_j \me^{-\mi\omega_\ell t}+\mathbf{p}_j^*\me^{\mi\omega_\ell t}\right)(\hat b_j^\dagger+\hat{b}_j),
\end{align}
with the classical amplitudes
\begin{align}
\mathbf{p}_j=Q_\mathrm{zpm}\left(\frac{\partial \alpha_\mathrm{m} (\hat{Q}_j)}{\partial  Q_j}\right)_{Q_0}\bE (\br_j^m),
\end{align}
and the free energy of the vibrational modes is given by $\hat{\mathcal H}_\mathrm{vib}=\sum_j\omega_\mathrm{vib} \hat{b}_j^\dagger \hat{b}_j^{\phantom{\dagger}}$. With this, the optomechanical Hamiltonian can be written as
\begin{align}
\hat{\Hm}_\mathrm{OM}=-\sum_j\sum_{\bk,\lambda}\mathcal{E}_\bk\left(\mathbf p_j \me^{-\mi\omega_\ell t}+\mathbf{p}_j^*\me^{\mi\omega_\ell t}\right)(\hat b_j^\dagger+\hat{b}_j^{\phantom\dagger})\left(\beps_{\bk,\lambda}\hat{a}_{\bk,\lambda}\me^{\mi\bk\cdot\br_j^m}+\beps_{\bk,\lambda}^*\hat{a}_{\bk,\lambda}^\dagger\me^{-\mi\bk\cdot\br_j^m}\right).
\end{align}
Under the assumption that all Raman dipoles are excited identically, i.e., $\mathbf p_j \equiv\mathbf p$ and making the RWA between Raman dipoles and electromagnetic modes which neglects simultaneous emission or absorption of two photons, one obtains
\begin{align}
\label{eq:hamomrwa}
\hat{\mathcal{H}}_\mathrm{OM}= \hbar\sum_j\sum_{\bk,\lambda}\left[g_{\bk,\lambda}^{\mathrm{OM}}\hat{a}_{\bk,\lambda}\me^{\mi\bk\cdot\br_j^\mathrm{m}}\me^{\mi\omega_\ell t}+g_{\bk,\lambda}^{\mathrm{OM},*}\hat{a}_{\bk,\lambda}^\dagger\me^{-\mi\bk\cdot\br_j^\mathrm{m}}\me^{-\mi\omega_\ell t}\right](\hat b_j^\dagger+\hat b_j^{\phantom\dagger}),
\end{align}
with coupling $g_{\bk,\lambda}^\mathrm{OM}=-\mathcal{E}_\bk (\beps_{\bk,\lambda}\cdot\mathbf{p}^*)/\hbar$.

Only considering the combinations in Eq.~\eqref{eq:hamomrwa} describing anti-Stokes processes (i.e., laser pump photons get up-converted in frequency aided by mechanical vibrations)
\begin{align}
\hat{\mathcal{H}}_\mathrm{OM}^\mathrm{AS}\approx \hbar\sum_j\sum_{\bk,\lambda}\left[g_{\bk,\lambda}^{\mathrm{OM}}\hat{a}_{\bk,\lambda}\hat b_j^\dagger\me^{\mi\bk\cdot\br_j^\mathrm{m}}\me^{\mi\omega_\ell t}+g_{\bk,\lambda}^{\mathrm{OM},*}\hat{a}_{\bk,\lambda}^\dagger\hat b_j \me^{-\mi\bk\cdot\br_j^\mathrm{m}}\me^{-\mi\omega_\ell t}\right],
\end{align}
leads to the equations of motion
\begin{subequations}
\begin{align}
\dot{\hat a}_{\bk,\lambda} &= -\mi\omega_\bk \hat{a}_{\bk,\lambda} -\mi \sum_j g_{\bk, \lambda}^{\bmu,*} \hat{A}_j\me^{-\mi\bk\cdot\br_j}-\mi\sum_j g_{\bk,\lambda}^{\mathrm{OM},*} \hat{b}_j\me^{-\mi\omega_\ell t}\me^{-\mi\bk\cdot\br_j^\mathrm{m}},\\
\dot{\hat{A}}_{j} &=-\mi\omega_0\hat{A}_j-\mi\sum_{\bk,\lambda}g_{\bk,\lambda}^{\bmu}\hat{a}_{\bk,\lambda}\me^{\mi\bk\cdot\br_j},\\
\dot{\hat{b}}_{j} &= -(\mi \omega_\mathrm{vib} +\Gamma_\mathrm{vib}/2)\hat{b}_j-\mi\sum_{\bk,\lambda}g_{\bk,\lambda}^{\mathrm{OM}}\hat{a}_{\bk,\lambda}\me^{\mi\omega_\ell t}\me^{\mi\bk\cdot\br_j^\mathrm{m}},
\end{align}
\end{subequations}
where we have additionally added a phenomenological vibrational damping rate $\Gamma_\mathrm{vib}$ in the equation of motion for $\hat b_j$.
To get rid of the explicit time dependence, one can go into a rotating frame by redefining $\hat{b}_j\to\hat{b}_j\me^{\mi\omega_\ell t}$
\begin{subequations}
\begin{align}
\dot{\hat a}_{\bk,\lambda} &= -\mi\omega_\bk \hat{a}_{\bk,\lambda} -\mi \sum_j g_{\bk, \lambda}^{\bmu,*} \hat{A}_j\me^{-\mi\bk\cdot\br_j}-\mi\sum_j g_{\bk,\lambda}^{\mathrm{OM},*} \hat{b}_j \me^{-\mi\bk\cdot\br_j^\mathrm{m}},\\
\dot{\hat{A}}_{j} &=-\mi\omega_0\hat{A}_j-\mi\sum_{\bk,\lambda}g_{\bk,\lambda}^{\bmu}\hat{a}_{\bk,\lambda}\me^{\mi\bk\cdot\br_j},\\
\dot{\hat{b}}_j  &= -\left[\mi (\omega_\mathrm{vib}+\omega_\ell )+\Gamma_\mathrm{vib}/2 \right]\hat{b}_j-\mi\sum_{\bk,\lambda}g_{\bk,\lambda}^{\mathrm{OM}}\hat{a}_{\bk,\lambda}\me^{\mi\bk\cdot\br_j^\mathrm{m}}.
\end{align}
\end{subequations}
This is completely analogous to the case discussed in the previous section and the solution for the MNP dipole amplitude in $\bq$ space can be expressed as
\begin{align}
\hat{A}_\bq (\omega)=\frac{\mi\bmu_0\cdot \hat{\mathbf{E}}_\bq^{\mathrm{in}, (-)}(\omega)}{\mi(\omega_0-\omega)+\Gamma_0^\mathrm{rad}/2-\mi\calS_\bq^{\bmu} (\omega)+\frac{\calS_\bq^{\mathrm{OM}}(\omega)^2}{\mi(\omega_\ell+\omega_\mathrm{vib}-\omega)+\Gamma_\mathrm{vib}/2+\gamma^\mathrm{rad}_{\mathrm{p},+}/2-\mi\calS_\bq^{\mathbf{p}}(\omega)}},
\end{align}
with the self-interaction of the Raman dipoles
\begin{align}
\calS_\bq^{\mathbf{p}}(\omega) =\frac{|\mathbf{p}|^2\omega^2}{ c^2\hbar\epsilon_0}\sum_{\Lambda \setminus \{0\} } \me^{-\mi \bq\cdot \br_\Lambda }\beps_{\mathbf{p}}\cdot \mathbf{G}(\br_\Lambda ,\omega)\cdot \beps_{\mathbf{p}},
\end{align}
and the ``optomechanical interaction'' between the Raman dipoles and the nanoparticle dipoles
\begin{align}
 \calS_\bq^{\mathrm{OM}}(\omega)  = \frac{\mu_{0} |\mathbf{p}|\omega^2}{c^2\hbar\epsilon_0}\sum_{\Lambda}\me^{-\mi \bq\cdot (\br_\Lambda+\br_\mathrm{m})}\beps_{\boldsymbol{\mu}}\cdot \mathbf{G}(\br_\Lambda+\br_\mathrm{m},\omega)\cdot \beps_{\mathbf{p}},
\end{align}
Now considering the Stokes combinations (pump photons get down-converted in frequency)
\begin{align}
\hat{\mathcal{H}}_\mathrm{OM}^\mathrm{S}\approx \hbar\sum_j\sum_{\bk,\lambda}\left[g_{\bk,\lambda}^{\mathrm{OM}}\hat{a}_{\bk,\lambda}\hat b_j^{\phantom\dagger}\me^{\mi\bk\cdot\br_j^\mathrm{m}}\me^{\mi\omega_\ell t}+g_{\bk,\lambda}^{\mathrm{OM},*}\hat{a}_{\bk,\lambda}^\dagger\hat b_j^\dagger \me^{-\mi\bk\cdot\br_j^\mathrm{m}}\me^{-\mi\omega_\ell t}\right],
\end{align}
the equations of motion are given by
\begin{subequations}
\begin{align}
\dot{\hat a}_{\bk,\lambda} &= -\mi\omega_\bk \hat{a}_{\bk,\lambda} -\mi \sum_j g_{\bk, \lambda}^{\bmu,*} \hat{A}_j\me^{-\mi\bk\cdot\br_j}-\mi\sum_j g_{\bk,\lambda}^{\mathrm{OM},*} \hat{b}_j^\dagger\me^{-\mi\omega_\ell t}\me^{-\mi\bk\cdot\br_j^\mathrm{m}},\\
\dot{\hat{A}}_{j} &=-\mi\omega_0\hat{A}_j-\mi\sum_{\bk,\lambda}g_{\bk,\lambda}^{\bmu}\hat{a}_{\bk,\lambda}\me^{\mi\bk\cdot\br_j},\\
\dot{\hat{b}}_{j}^\dagger &= (\mi\omega_\mathrm{vib} -\Gamma_\mathrm{vib}/2)\hat{b}_j^\dagger+\mi\sum_{\bk,\lambda}g_{\bk,\lambda}^{\mathrm{OM}}\hat{a}_{\bk,\lambda}\me^{\mi\omega_\ell t}\me^{\mi\bk\cdot\br_j^\mathrm{m}},
\end{align}
\end{subequations}
In a rotating frame $\hat b_j^\dagger\to\hat b_j^\dagger \me^{-\mi\omega_\ell t}$, we obtain
\begin{subequations}
\begin{align}
\dot{\hat a}_{\bk,\lambda} &= -\mi\omega_\bk \hat{a}_{\bk,\lambda} -\mi \sum_j g_{\bk, \lambda}^{\bmu,*} \hat{A}_j\me^{-\mi\bk\cdot\br_j}-\mi\sum_j g_{\bk,\lambda}^{\mathrm{OM},*} \hat{b}_j^\dagger\me^{-\mi\bk\cdot\br_j^\mathrm{m}},\\
\dot{\hat{A}}_{j} &=-\mi\omega_0\hat{A}_j-\mi\sum_{\bk,\lambda}g_{\bk,\lambda}^{\bmu}\hat{a}_{\bk,\lambda}\me^{\mi\bk\cdot\br_j},\\
\dot{\hat{b}}_{j}^\dagger &= \left[\mi(\omega_\mathrm{vib}-\omega_\ell)-\Gamma_\mathrm{vib}\rev{/2}\right] \hat{b}_j^\dagger+\mi\sum_{\bk,\lambda}g_{\bk,\lambda}^{\mathrm{OM}}\hat{a}_{\bk,\lambda}\me^{\mi\bk\cdot\br_j^\mathrm{m}},
\end{align}
\end{subequations}
and the solution for the MNP dipole amplitudes is given by
\begin{align}
\hat{A}_\bq (\omega)=\frac{\mi\bmu_0\cdot \hat{\mathbf{E}}_\bq^{\mathrm{in}, (-)}(\omega)}{\mi(\omega_0-\omega)+\Gamma_0^\mathrm{rad}/2-\mi\calS_\bq^{\bmu} (\omega)-\frac{\calS_\bq^{\mathrm{OM}}(\omega)^2}{\mi(\omega_\ell-\omega_\mathrm{vib}-\omega)+\Gamma_\mathrm{vib}/2-\gamma^\mathrm{rad}_{\mathrm{p},-}/2+\mi\calS_\bq^{\mathbf{p}}(\omega)}}.
\end{align}
The decay rates for the Raman dipoles induced by the optomechanical interaction for the anti-Stokes/Stokes sidebands are given by $\gamma_{\mathrm{p},\pm}^\mathrm{rad}=|\mathbf p |^2(\omega_\ell\pm\omega_\mathrm{vib})^3/(3\pi c^3\hbar\epsilon_0)$. It should be noted that the RWA for the interaction between vibrational and photonic modes has to be applied with caution, particularly in the blue-detuned regime (see discussion in main text). This is because the broad plasmonic background associated with the SLR can overlap spectrally with the anti-Stokes sideband at higher energies. In such cases, neglecting counter-rotating terms may lead to a wrong prediction of the impact of the optomechanical coupling onto the optical response. If significant overlap is present, a full non-RWA treatment is warranted to correctly describe the optical response. 

\rev{We also note that fluctuations of molecular positions and phases within the unit cells can be incorporated by replacing $\br_\mathrm{m}\to\br_\mathrm{m}+\delta\br_{\mathrm{m},\Lambda}$, and $\mathbf{p}\to \mathbf{p}\me^{\mi\Phi_\Lambda}$, which modifies the sum to 
\begin{align}
\calS_\bq^{\mathrm{OM}}(\omega) = \frac{\mu_{0} |\mathbf{p}|\omega^2}{c^2\hbar\epsilon_0}\sum_{\Lambda}\me^{-\mi \bq\cdot (\br_\Lambda+\br_\mathrm{m}+\delta\br_{\mathrm{m},\Lambda})}\me^{\mi\Phi_\Lambda}\beps_{\boldsymbol{\mu}}\cdot \mathbf{G}(\br_\Lambda+\br_\mathrm{m}+\delta\br_{\mathrm{m},\Lambda},\omega)\cdot \beps_{\mathbf{p}},
\end{align}
generally leading to a renormalization of the lattice sum and therefore the optomechanical coupling. 
Since for most of the results presented in the main text, we focus on modes with $|\bq|\approx 0$, whose wavelength is much larger than the lattice constant. In this long-wavelength limit, the factors $\me^{-\mi\bq\cdot\delta\br_{\mathrm{m},\Lambda}}$ are close to unity, such that fluctuations of the molecular positions (and similarly of the phases) are strongly suppressed in the lattice sum. As a result, these non-idealities produce only minor corrections and do not modify the qualitative behaviour discussed in the main text.}

\section*{S5. Nonlinear polarizability of excitonic SLRs for static populations}

Here, we derive the polarizability of the transition dipoles for an excitonic array. The interaction of the excitonic array with the electromagnetic field is described by
\begin{align}
\hat{\calH}_\text{int}=\hbar\sum_{j,\nu<\nu'}\sum_{\bk,\lambda} g_{\bk, \lambda}^{\nu\nu'}   \left(\hat a_{\bk,\lambda}^{\dagger} \me^{-\mi \bk\cdot\br_j}\hat \sigma_{j}^{\nu\nu'} +\mathrm{H.c.}\right),
\end{align}
where $\hat\sigma_j^{\nu\nu'}=\ket{\nu}_j\bra{\nu'}_j$ describes the lowering operator between transitions $\nu$ and $\nu'$ for a given molecule $j$ and the coupling with the electromagnetic field is given by $g_{\bk, \lambda}^{\nu\nu'} =-\mi \mathcal{E}_\bk (\beps_{\bk,\lambda}\cdot \bmu_{\nu\nu'})/\hbar$. 

One can now follow the same procedure as detailed in Sec.~S1 in eliminating the photon modes to derive an effective equation for the matter part in momentum space. The dipole operator of the $\nu\nu'$-transition can be expressed as 
\begin{align}
-\mi\omega\hat\sigma_{\bq}^{\nu\nu'}(\omega)&=-\left(\mi\omega_{\nu\nu'}+\frac{\Gamma_{\nu\nu'}}{2}\right)\hat{\sigma}_\bq^{\nu\nu'}(\omega)+\mi \calS_{\bq}^{\nu\nu'}(\omega)\hat{\sigma}_{\bq}^{\nu\nu'}(\omega) p_{\mathrm{inv}}^{\nu\nu'}-p_\mathrm{inv}^{\nu\nu'}\hat{\sigma}_\bq^{\nu\nu', \mathrm{in}}(\omega),
\end{align}
where the decay rates of the different transitions are given by $\Gamma_{\nu\nu'}^\mathrm{rad}=|\boldsymbol{\mu}_{\nu\nu'}|^2\omega_{\nu\nu'}^3/(3\pi c^3\hbar\epsilon_0)$. Here, the collective input noise term affecting the electronic transitions is defined as
\begin{align}
\hat{\sigma}_\bq^{\nu\nu', \mathrm{in}}(\omega)=-\mi\sum_\lambda g_{\bq,\lambda}^{\nu\nu'}\hat a_{\bq,\lambda} (0) \me^{-\mi\omega_\bq t}.
\end{align}
Importantly, here we have made the assumption of equal and constant population in all sublevels $p_{j,\mathrm{inv}}^{\nu\nu'}=\expval{\hat{\sigma}_{j, \nu\nu'}^z}\equiv p_\mathrm{inv}^{\nu\nu'}$. Furthermore, we have neglected interactions among the different subtransitions by assuming that the frequencies of the different electronic transitions are very off-resonant.
The lattice sum of the $\nu\nu'$ transition dipoles is given by
\begin{align}
\calS_{\bq}^{\nu\nu'}(\omega)=3\pi \Gamma_{\nu\nu'}^{\mathrm{rad}} c \,\frac{\omega^2}{\omega_{\nu\nu'}^3}\left(\sum_{\Lambda \setminus \{0\} } \me^{-\mi \bq \cdot\br_\Lambda} \beps_{\boldsymbol \mu_{\nu\nu'}}\cdot \mathbf G (\br_\Lambda,\omega) \cdot \beps_{\boldsymbol\mu_{\nu\nu'}}\right).
\end{align}
The solution for the dipole operator of the $\nu\nu'$ transition in Fourier space then expresses as
\begin{align}
\hat\sigma_{\bq}^{\nu\nu'}(\omega)=\frac{-p_\mathrm{inv}^{\nu\nu'}\hat\sigma_\bq^{\nu\nu',\mathrm{in}}(\omega)}{\iu\left(\omega_{\nu\nu'}-\omega\right)+\Gamma_{\nu\nu'}^\mathrm{rad}/2+\iu p_\mathrm{inv}^{\nu\nu'}\mathcal{S}_{\bq}^{\nu\nu'}(\omega)}.
\end{align}
\revise{This corresponds to the linear response result, as will be shown in the following section S5.}

\section*{S6. Pump-probe spectroscopy of nonlinear SLR dynamics}
\label{sec:supps5}

In this section, we show  how to explicitly include population dynamics into our treatment. 
Neglecting cross-coupling between different dipole transitions, the equations of motion for the coherences express in time domain as
\begin{subequations}
\begin{align}
\dot{\hat{\sigma}}_{j}^{12}(t)&=-\mi\omega_{12}\hat{\sigma}_j^{12}-\hat{\sigma}_j^{12,\mathrm{in}}(t)\hat{p}_{\mathrm{inv}, j}^{12}(t)+\hat{p}_{\mathrm{inv}, j}^{12}(t)\sum_{\bk, \lambda} |g_{\bk, \lambda}^{12}|^2\sum_{j'}\int_0^t\td t' \me^{-\mi\omega_\bk (t-t')}\hat{\sigma}_{j'}^{12}(t')\me^{\mi \bk\cdot(\br_{j}-\br_{j'})},\\
\dot{\hat{\sigma}}_{j}^{23}(t)&=-\mi\omega_{23}\hat{\sigma}_j^{23}-\hat{\sigma}_j^{23,\mathrm{in}}(t)\hat{p}_{\mathrm{inv}, j}^{23}(t)+\hat{p}_{\mathrm{inv}, j}^{23}(t)\sum_{\bk, \lambda} |g_{\bk, \lambda}^{23}|^2\sum_{j'}\int_0^t\td t' \me^{-\mi\omega_\bk (t-t')}\hat{\sigma}_{j'}^{23}(t')\me^{\mi \bk\cdot(\br_{j}-\br_{j'})},
\end{align}
\end{subequations}
while the equation of motion for the population of the $\ket{2}$ state $\ket{2}_j\bra{2}_j=\hat{\sigma}_j^{12,\dagger}\hat{\sigma}_j^{12}$ is given by
\begin{align}
\partial_t [\hat{\sigma}_j^{12,\dagger}\hat{\sigma}_j^{12}](t)&=\hat{\sigma}_j^{12,\mathrm{in}}(t) \hat{\sigma}_{ j}^{12, \dagger}(t)+\hat{\sigma}_j^{12,\mathrm{in}}(t)^\dagger\hat{\sigma}_{ j}^{12}(t)-\left(\hat{\s}_j ^{12, \dagger} (t)\sum_{\bk, \lambda} |g_{\bk, \lambda}^{12}|^2\sum_{j'}\int_0^t \td t' \me^{-\mi\omega_\bk (t-t')}\hat{\s}_{j'}^{12} (t')\me^{\mi \bk\cdot (\br_j-\br_{j'})}+\mathrm{H.c}\right)+\hdots,
\end{align}
where the $\hdots$ denotes other terms which become important beyond third order for our perturbative treatment below (third-order coherences feeding back into populations) and which we therefore drop for the ease of notation. 

\begin{figure*}[h!]
  \centering
  \includegraphics[width=0.3\textwidth]{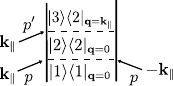}
		\caption{\revise{\textbf{Feynman diagram of third-order process.} The pump field ($p$) generates zero-momentum population via two interactions with opposite phase. The probe field ($p'$) then leads to the generation of coherence between the $\ket{2}$ and $\ket{3}$ states.}}
	\label{figS2}
\end{figure*}

Generally, this leads to an analytically intractable set of equations involving convolutions in the time domain and space. 
A simplification can be carried out by considering a perturbative expansion in terms of input amplitudes \cite{mukamel1995principles, reitz2025nonlinear}. 
To this end, we consider input pulses
\begin{align}
\langle \hat{\sigma}_j^{12,\mathrm{in}}(t) \rangle &= \eta_p f_p(t) \, \me^{-\mathrm{i} \omega_p t}\me^{\mi\bk_{\parallel,p}\cdot\br_j}, \qquad
\langle \hat{\sigma}^{23,\mathrm{in}}_j(t) \rangle = \eta_{p'} f_{p'}(t-\tau_\Delta) \, \me^{-\mathrm{i} \omega_{p'} t}\me^{\mi\bk_{\parallel,p'}\cdot\br_j},
\end{align}
where $\eta_{p,p'}$ describes the pulse amplitudes (considered real here), $f_{p,p'}(t)$ the temporal envelope (assumed Gaussian), $\tau_\Delta$ is the delay time between the pulses, and we assume incidence of pump and probe pulses at \rev{wavevectors $\bk_{\parallel, p/p'}$}.
We then go to momentum space for both coherences and population operators
\begin{align}
\hat{\s}_j^{\nu\nu'}=\frac{1}{\sqrt{M}}\sum_\bq \hat{\s}_\bq^{\nu\nu'}\me^{\mi\bq\cdot\br_j},\quad [\hat{\s}^{\nu\nu',\dagger}\hat{\s}^{\nu\nu'}]_j=\frac{1}{\sqrt{M}}\sum_\bq[\hat{\s}^{\nu\nu',\dagger}\hat{\s}^{\nu\nu'}]_\bq\me^{\mi\bq\cdot\br_j},
\end{align}
and expand all operators in terms of their contribution in the input fields as
\begin{align}
\langle\hat{O}_\bq\rangle = \sum_{n,m=0}^\infty \eta_p^n \eta_{p'}^m O_\bq^{(n)(m)}.
\end{align}
At each order, a closed set of equations for the expansion coefficients $O_\bq^{(n)(m)}$ can then be derived \rev{under the mean-field assumption $\expval{\hat{\s}_j \hat{\s}_{j'}} = \expval{\hat{\s}_j}\expval{\hat{\s}_{j'}}=\s_j\s_{j'}$ for $j \neq j'$ (and similarly for $\expval{\hat{p}_{\mathrm{inv},j}\hat{\s}_{j'}}$)}. In particular, in the following we will consider the modification of the probe response due to the second-order populations generated by the pump, corresponding to order $(2)(1)$ [see Feynman diagrams in Fig.~\ref{figS2}].\\

\paragraph{First order in pump: $(1)(0)$.}

At first order in the pump field, only coherence between $\ket{1}$ and $\ket{2}$ is created and populations are static:
\begin{align}
-\mi\omega\s_{\bq}^{12}(\omega) ^{(1)(0)}
&= -\left( \mi\omega_{12} +\frac{\Gamma_{12}^\mathrm{rad}}{2} \right)\s_{\bq}^{12}(\omega) ^{(1)(0)}
 - \mi\calS_{\bq}^{12}(\omega)\s_{\bq}^{12}(\omega) ^{(1)(0)} p_\mathrm{inv}^{12}(0)
 - \delta_{\bq, \bk_{\parallel,p}} f_{p}(\omega-\omega_p) p_{\mathrm{inv}}^{12}(0),
\end{align}
where the populations are fixed by their initial conditions $p_{\mathrm{inv}}^{12}(0)=p_{\mathrm{inv},j}^{12}(0)=[p_{\mathrm{inv},j}^{12}]^{(0)(0)}$, and $f_{p}(\omega)$ denotes the Fourier transform of $f_{p}(t)$. For initial condition $p_{\mathrm{inv},j}^{12}(0)=-1$ (all population initially in the ground state), we obtain
\begin{align}
-\mi\omega\s_{\bq}^{12}(\omega) ^{(1)(0)}
&= -\left( \mi\omega_{12} + \frac{\Gamma_{12}^\mathrm{rad}}{2} \right)\s_{\bq}^{12}(\omega) ^{(1)(0)}
 + \mi\calS_{\bq}^{12}(\omega)\s_{\bq}^{12}(\omega) ^{(1)(0)}
 + \delta_{\bq, \bk_{\parallel,p}} f_{p}(\omega-\omega_p).
\end{align}
Therefore, at linear order only the momentum mode $\bq=\bk_{ \parallel,p}$ (\rev{we restrict all wavevectors to the $1$st Brillouin zone}) is driven (and $\bq=-\bk_{ \parallel,p}$ for the complex conjugate coherence $\s_{\bq}^{12,*}$).\\

\paragraph{Second order in pump: $(2)(0)$.}

\rev{At second order, the coherences created at first order can create populations in the $\ket{2}$ state 
\begin{align}
-\mi\omega \left[ \s^{12,\dagger}\s^{12} \right]_{\bq} (\omega)^{(2)(0)}
&= -\Gamma_{12}^\mathrm{rad}\left[ \s^{12,\dagger}\s^{12} \right]_{\bq} (\omega)^{(2)(0)}-2\mathrm{Im}\sum_{\mathbf{p}}\left[\s_{\mathbf{p}}^{12,*}(\omega)^{(1)(0)}\ast\calS_{\mathbf{q}-\mathbf{p}}^{12}\s_{\mathbf{q}-\mathbf{p}}^{12}(\omega)^{(1)(0)}\right]\\\nonumber
&+\left[\s_{\bq-\bk_{\parallel,p}}^{12,*}(\omega)^{(1)(0)}\ast f(\omega-\omega_p)\right]+\left[\s_{\bq+\bk_{\parallel,p}}^{12}(\omega)^{(1)(0)}\ast f(\omega-\omega_p)\right],
\end{align}
generally leading to convolutions in both momentum and frequency space. Here, an asterisk in the exponent describes complex conjugation and an asterisk between two symbols denotes convolution. Making use of the fact that the only momentum mode driven for the coherence at linear order is $\bk_{ \parallel,p}$ (and the opposite sign for the complex conjugate), we can see that only the zero-momentum mode is driven for the population at second order (see Fig.~\ref{figS2})
\begin{align}
-\mi\omega \left[ \s^{12,\dagger}\s^{12} \right]_{\bq=0} (\omega)^{(2)(0)}
&= -\Gamma_{12}^\mathrm{rad}\left[ \s^{12,\dagger}\s^{12} \right]_{\bq} (\omega)^{(2)(0)} -2\mathrm{Im}\left[\s_{-\bk_{ \parallel,p}}^{12,*}(\omega)^{(1)(0)}\ast\calS_{\bk_{ \parallel,p}}^{12}\s_{\bk_{ \parallel,p}}^{12}(\omega)^{(1)(0)}\right]\\\nonumber
&+\left[\s_{-\bk_{\parallel,p}}^{12,*}(\omega)^{(1)(0)}\ast f(\omega-\omega_p)\right]+\left[\s_{\bk_{\parallel,p}}^{12}(\omega)^{(1)(0)}\ast f(\omega-\omega_p)\right].
\end{align}}\\

\paragraph{Second order in pump, first order in probe: $(2)(1)$.}

These second-order populations can now act as the generator of third-order coherences $\s_{\bq}^{23}(\omega) ^{(2)(1)}$ (by creating a non-zero inversion in the $2\!\leftrightarrow\!3$ transition) via application of the probe field $p'$:
\begin{align}
-\mi\omega\s_{\bq}^{23}(\omega) ^{(2)(1)}
&= -\left( \mi\omega_{23} + \frac{\Gamma_{23}^\mathrm{rad} }{2} \right)\s_{\bq}^{12}(\omega) ^{(2)(1)}
 + \left[ f_{p'}(\omega-\omega_{p'}) \ast \left( \s^{12,\dagger}\s^{12} \right)_{\bq-\bk_{\parallel,p'}} (\omega)^{(2)(0)} \right](\omega)
 + \mi\mathcal{S}_\bq^{23}(\omega)\s_\bq^{23}(\omega)^{(2)(1)}.
\end{align}
Therefore, the momentum mode that is driven for the coherence at third order is $\bq=\bk_{\parallel,p'}$, since only the population with zero quasi-momentum is contributing.

\end{document}